\documentclass[final,5p,times]{elsarticle}

\usepackage{amssymb}

\usepackage{float}
\usepackage{subfig}
\usepackage[export]{adjustbox}
\usepackage{placeins}
\usepackage{booktabs}
\newcommand{\head}[1]{\textnormal{\textbf{#1}}}
\usepackage{amsmath,bm}
\usepackage{pdflscape}
\usepackage{url}
\usepackage{textgreek}
\usepackage[final]{microtype}

\usepackage[usenames]{color}

\usepackage{adjustbox}

\newcommand{\beginsupplement}{%
        \setcounter{table}{0}
        \renewcommand{\thetable}{S\arabic{table}}%
        \setcounter{figure}{0}
        \renewcommand{\thefigure}{S\arabic{figure}}%
        \setcounter{section}{0}
        \renewcommand{\thesection}{S\arabic{section}}
     }

\journal{Additive Manufacturing}

\begin{document}

\begin{frontmatter}

\title{Spatial Mapping of Powder Layer Density for Metal Additive Manufacturing \\ via X-ray Microscopy}

\author[MIT]{Ryan~W.~Penny\corref{cor1}}
\ead{rpenny@mit.edu}
\author[TUM]{Patrick~M.~Praegla}
\author[TUM]{Marvin~Ochsenius}
\author[MIT]{Daniel~Oropeza}
\author[TUM]{Christoph~Meier}
\author[TUM]{Wolfgang~A.~Wall}
\author[MIT]{A.~John~Hart\corref{cor1}}
\ead{ajhart@mit.edu}

\address[MIT]{Department of Mechanical Engineering, Massachusetts Institute of Technology, 77 Massachusetts Avenue, Cambridge, 02139, MA, USA}
\address[TUM]{Institute for Computational Mechanics, Technical University of Munich, Boltzmannstra{\ss}e 15, Garching b. M{\"u}nchen, Germany}

\cortext[cor1]{Corresponding author}

\begin{abstract}

Uniform powder spreading is a requisite for creating consistent, high-quality components via powder bed additive manufacturing (AM), wherein layer density and uniformity are complex functions of powder characteristics, spreading kinematics, and mechanical boundary conditions.  High spatial variation in particle packing density, driven by the stochastic nature of the spreading process, impedes optical interrogation of these layer attributes.  Thus, we present transmission X-ray imaging as a method for directly mapping the effective depth of powder layers at process-relevant scale and resolution. Specifically, we study layers of nominal 50-250~\textmu m thickness, created by spreading a selection of commercially obtained Ti-6Al-4V, 316 SS, and Al-10Si-Mg powders into precision-depth templates. We find that powder layer packing fraction may be predicted from a combination of the relative thickness of the layer as compared to mean particle size, and flowability assessed by macroscale powder angle of repose.  Power spectral density analysis is introduced as a tool for quantification of defect severity as a function of morphology, and enables separate consideration of layer uniformity and sparsity.  Finally, spreading is studied using multi-layer templates, providing insight into how particles interact with both previously deposited material and abrupt changes in boundary condition.  Experimental results are additionally compared to a purpose-built discrete element method (DEM) powder spreading simulation framework, clarifying the competing role of adhesive and gravitational forces in layer uniformity and density, as well as particle motion within the powder bed during spreading.

\end{abstract}

\begin{keyword}
Additive Manufacturing \sep Powder Spreading \sep X-ray \sep Defects \sep DEM
\end{keyword}

\end{frontmatter}

\section{Introduction}
\label{sec:intro}

Powder spreading is critical to stable processing and to building high quality components in powder bed additive manufacturing (AM), as the powder layer influences process conditions and component properties in many ways~\cite{Vock2019, Ziaee2019}.  For example, laser powder bed fusion (LPBF) is a common metal powder bed fusion (PBF) AM process relying on laser-delivered energy to melt and fuse powdered feedstock into a monolithic component.  Subsurface scattering of the laser primarily drives absorption of energy; this is a volumetric process where deeper and more densely packed powder layers provide more potential photon-particle interactions for absorption of laser beam energy ~\cite{Gusarov2005ModellingTreatment, Boley2015CalculationManufacturing, Boley2016MetalExperiment}.  Similarly, melt pool fluid dynamics are influenced by powder properties and packing behavior. Sub-optimally thick powder layers have been correlated with decreased melt pool stability, increased surface roughness, and defects in components produced by LPBF~\cite{Lee2015, Qiu2015OnMelting, Lee2015, Mindt2016, Escano2018}.

Powder feedstock used in LPBF is typically created via gas or plasma atomization, as the particles produced are spherical and therefore spread more predictably than irregular particle shapes (e.g., produced by water atomization) ~\cite{Boley2016MetalExperiment, Tan2017AnProcess}.  Particle size distributions used in LPBF typically range between $10$ and $60$~\textmu m, achieving a compromise between small layer thickness, which enables high resolution processing, and favorable spreading mechanics~\cite{Mindt2016, Vock2019, Brandt2017, Sutton2016}.  Specifically, to avoid defects due to streaking by large particles, the ideal thickness of spread powder layers may be estimated as no less than $1.5\times D_{90}$, where $D_{90}$ is the 90\textsuperscript{th} percentile of particle diameters in the size distribution~\cite{Spierings2011, Mindt2016}.  Layers that are thinner than desirable are further prone to bridging, wherein particles at the bottom of the layer fail to settle to a dense configuration~\cite{Chen2017}.  Moreover, cohesive (i.e., van der Waals) forces between powder particles become comparable to gravitational forces for fine powder particles, causing powder clumping and corresponding poor flowability~\cite{Yablokova2015RheologicalImplants, Chen2017, Meier2019CriticalManufacturing, Meier2019ModelingSimulations}.

Even in cases where cohesive forces do not dominate powder behavior, achieving uniform spreading can be challenging. For instance, Wischeropp and coworkers report 1\textsigma {} standard deviations spanning 6-22\% about an average packing fraction of $\approx 50$\% within powder specimens retrieved from a commercial LPBF machine~\cite{Wischeropp2019}.  Nonuniformity is ascribed primarily to random fluctuation of layer thickness.  Moreover, consolidation of material causes layer thickness to evolve throughout the printing process~\cite{Mindt2016}.  Using typical LPBF process parameters of 50\% packing fraction and 50~\textmu m nominal layer thickness for illustration, the first powder layer in a build will densify into a 25~\textmu m thick solid layer, and therefore increase the thickness of the second powder layer to 75~\textmu m (i.e., 50~\textmu m nominal layer thickness plus 25~\textmu m from vertical consolidation of the prior layer).  Powder layer thickness approaches the nominal layer thickness divided by the packing fraction as the build progresses, such that the thickness of the fused layers equals the nominal layer thickness~\cite{Mindt2016}.  Thus, layer thickness, density, and defect probability are all highly variable.

Electron beam melting (EBM) and binder jetting (BJ) are other common powder bed AM techniques, and likewise rely on uniform powder spreading to achieve consistent processing and component performance.  EBM is analogous to LPBF, in that a focused and scanned electron beam is used to locally melt layers of metal powder. A larger particle size distribution than LPBF (typically $45-106$ \textmu m) is necessitated by the physics of EBM, as the increased particle mass (gravitational force) counters electrostatic forces arising from particle charging~\cite{Gong2014, Korner2016}.  BJ uses a binder, deposited by inkjet printing layer-by-layer, to adhere particles prior to post-processing steps comprising debinding, sintering, and, in some cases, infiltration.  As the feedstock is never fully melted, high packing density of each layer directly drives both high component density and mechanical performance~\cite{Ziaee2019}, and use of finer powders can reduce material cost and sintering time.  As such, strategies to achieve uniform and dense layers in BJ rely upon finer powders (mean particle diameter of $\approx 10-30$ \textmu m) than typical in LPBF, and sometimes incorporate bimodal size distributions~\cite{Budding2013,Bai2017, Ziaee2019}.  However, the high cohesive forces and poor flowability associated with these fine powders necessitates alternative spreading strategies~\cite{Ziaee2019}, such as use of roller-based compaction~\cite{Shanjani2008, Rishmawi2018}, mechanical vibration~\cite{Jimenez2019}, and fluidized-hopper-based powder deposition~\cite{Pruitt1991}.

Many researchers have explored the influence of the spreading tool on layer uniformity.  In addition to adapting to underlying component topography, for example, compliant spreading (recoating) blades increase contact area with flowing powder particles and thereby improve uniformity and reduce surface roughness of powder layers~\cite{Le2021}. Specifically, work by Snow and coworkers~\cite{Snow2019} indicates that an elastomeric (silicone) blade is specifically beneficial when spreading cohesive powders.  Likewise, discrete element method (DEM) simulations of non-spherical powders show that counter-rotating rollers produce more uniform layers than a rectangular blade with $90^\circ$ corners~\cite{Haeri2017}; however, a rigid blade may be made nearly as effective by changing corner geometry~\cite{Haeri2017_2}. Overall, spreading with a (counter-rotating) roller along with precise metering of the powder is effective for finer and more cohesive powders~\cite{Ziaee2019}. Traverse speed of the spreading tool is also important, and excessively high speeds can cause sparse powder deposition~\cite{Chen2017, Snow2019, Meier2019CriticalManufacturing,  Lee2020}.  Finally, damage to the spreading tool, distortion of the underlying component due to thermal stress in LPBF, and mechanical vibrations of the spreading apparatus may also manifest as layer nonuniformities~\cite{Vock2019, Hendriks2019}, which in turn influence the laser-material interaction and melt pool characteristics.

Optical imaging techniques are often used to study powder spreading for PBF AM (e.g.,~\cite{Craeghs2011_2, Kleszczynski2012, Hendriks2019}).  As illustrated by the work of Snow and coworkers, the probability of bare regions in a powder layer is non-linearly related to, inter alia, angle of repose, recoating speed, and blade material~\cite{Snow2019}.  However, difficulties with this approach of optical imaging are described by Phuc and Seita~\cite{TanPhuc2019AManufacturing}; namely, spatial resolution is limited by detector resolution, diffraction, perspective distortion, and limited depth of field. Alternatively, Phuc and Seita use the linear CCD sensor from a flatbed scanner to image powder beds at $5$~\textmu m spatial resolution, and demonstrate measurement of surface topography and identification of defects therefrom. While these optical imaging techniques are sensitive to variations in layer height and presence of bare regions, and have a clear path towards industrial application, they cannot sense variation in the volume of material deposited or equivalently the local packing density \cite{Ali2018OnProcesses, TanPhuc2019AManufacturing}.

X-ray imaging methods provide an alternative to optical sensing.  Side-on imaging is described by Escano and coworkers, who use a synchrotron X-ray imaging system to record video while spreading a narrow ($5$~mm width in beam direction) powder bed~\cite{Escano2018}.  The authors observe that particle clusters substantially disrupt smooth powder flow, and that the upper surface of a layer of fine powder is considerably smoother ($D_{50} = 23$~\textmu m, $R_a = 20$~\textmu m) than a layer fabricated with coarse powders ($D_{50} = 67$~\textmu m, $R_a = 37$~\textmu m).  Alternatively, Ali et al.\ have developed a method for preparing powder beds for ex-situ CT imaging, using a polymer binder to fuse small (mm$^3$ scale) representative regions from a powder bed~\cite{Ali2018OnProcesses}.  Results indicate that packing density decreases with recoater travel distance due to an evolving diameter distribution, arising from preferential deposition of relatively small, gap-filling particles.  However, powder particle motion upon injection of binders into a powder bed is a well known phenomenon~\cite{Sachs1993, Parab2019}; thus, the samples imaged may not provide a highly accurate representation of the as-spread layer.

Therefore, to advance understanding of powder bed AM, we emphasize the need for an experimental method that directly resolves the volume of powder deposited under industrially-relevant spreading conditions and size scales. Accordingly, we present an X-ray microscopy technique illustrated in Fig.~\ref{fig:Intro}.  Specimen layers are created by spreading powder into templates with tightly controlled dimensions, as illustrated in Figs.~\ref{fig:Intro}c and~\ref{fig:Intro}d.  X-ray imaging of these layers enables calculation of the effective depth of powder deposited, defined as the total volume of material deposited in a region (see graphical definition in Fig.~\ref{fig:MethodX-ray}a), as demonstrated by the corresponding effective depth map shown in Fig.~\ref{fig:Intro}e.  In the sections below, the instrument and technique for measuring powder layers are first described, followed by interpretation of layer density and uniformity from effective depth data.  Random and systematic non-uniformities are quantified, including variance due to streaking defects and changes in deposition due to powder flow obstruction at template edges evident in the exemplary powder layer.

\section{Methods}
\label{sec:methods}

\subsection{X-Ray Microscopy Apparatus}
\begin{figure}[ht]
\begin{center}
	{
	\includegraphics[trim = {.75in 1.3in 4.4in 1.45in}, clip, scale=1, keepaspectratio=true]{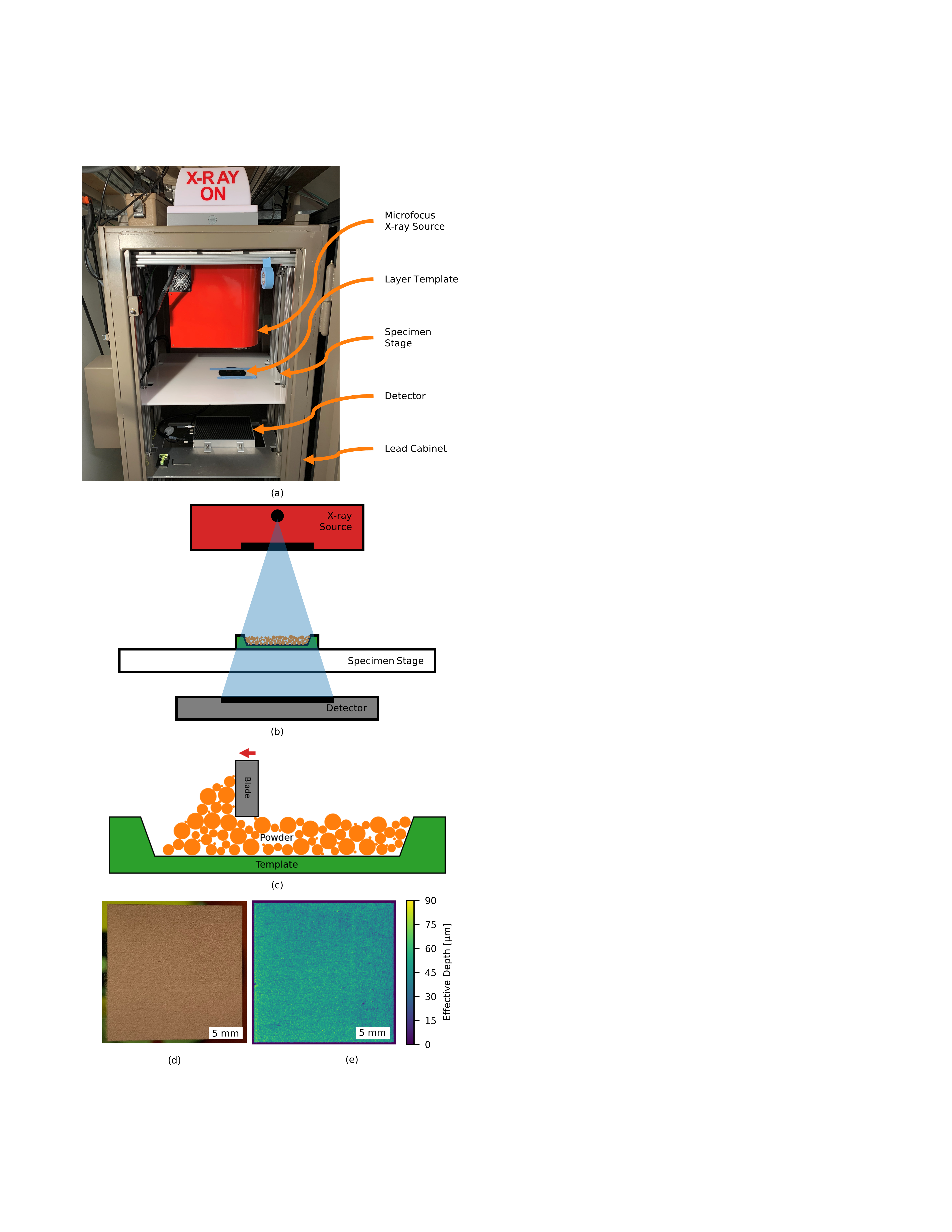}
	}
\end{center}
\vspace{-11pt}
\caption{Method for X-ray imaging of powder layers.  (a) X-ray microscope for measurement of powder layer transmission. (b) Schematic illustration of X-ray beam path (blue).  (c) Illustration of powder layer creation using a blade and template.  (d) Optical image of a typical powder layer.  (e) Typical measurement of powder layer effective depth.}
\label{fig:Intro}
\end{figure}

Figure~\ref{fig:Intro}a shows the custom-built X-ray microscope system used herein.  The instrument centers on a microfocus X-ray source (Hamamatsu L12161-07), that features considerable adjustability in tube output ($0-150$~kV, $0-500$~\textmu A) and emission spot size ($5-50$~\textmu m). As shown schematically in Fig.~\ref{fig:Intro}b, the source generates a beam passing downwards, through a polyoxymethylene (acetal) sample stage, to an imaging X-ray detector (Varex 1207 NDT, $75$~\textmu m pixel size).  The source, specimen stage, and detector are disposed on an adjustable metal frame within a custom-built lead cabinet (Hopewell Designs), thereby enabling adjustable geometric magnification (3X in the configuration shown) and efficient beam use. Radiation-protection safeguards comprise interlocks on cabinet doors, mandatory Geiger counter surveys before experiments are performed, and film dosimeters to detect any departure from baseline radiation levels.

\subsubsection{Templates for Powder Layers}

\begin{figure*}[ht]
\begin{center}
	{
	\includegraphics[trim = {1.65in 4.75in .67in 1.9in}, clip, scale=1, keepaspectratio=true]{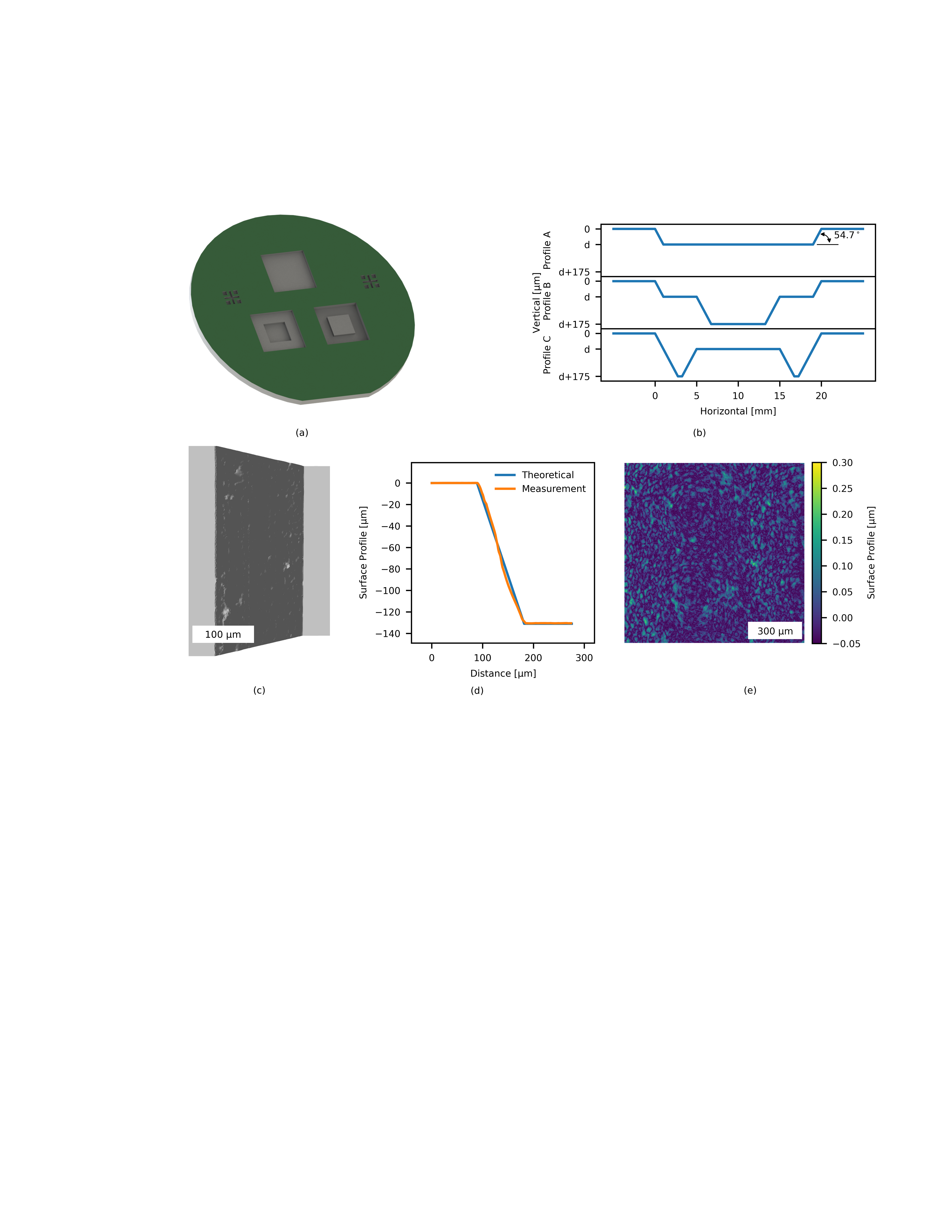}
	}
\end{center}
\vspace{-11pt}
\caption{Design and qualification of powder layer templates. (a) CAD rendering depicting three template designs on one wafer.  (b) Nominal template profiles.  (c) Confocal microscope image of a template edge.  (d) Profile of a nominal 131~\textmu m depth template edge.  (e) Typical profile of a template bottom surface.}
\label{fig:Templates}
\end{figure*}

Model powder layers are created by spreading powder into uniform-depth templates, which are fabricated in silicon wafers per Fig.~\ref{fig:Templates}, and are affixed to the specimen stage as shown in Fig.~\ref{fig:Intro}a. Silicon wafers are chosen because of their flatness and low surface roughness, as well as established microfabrication (i.e., chemical etching) techniques.  The precision enabled thereby permits control of template dimensions to sub-micron tolerances; thus, powder flow is not disturbed by geometric errors of the template and fluctuations in effective depth may be directly ascribed to the spreading process itself.  Layer templates are batch fabricated on a single wafer as shown in the rendering of Fig.~\ref{fig:Templates}a, with cross-sections dimensioned in Fig.~\ref{fig:Templates}b.  To study the influence of layer height on spreading and layer characteristics, templates are fabricated in a range of depth values ($d$ in the profile illustrations), logarithmically spaced at 50, 69, 95, 131, 181, and 250~\textmu m.  Templates with a uniform, flat bottom (Profile A) are fabricated, as well as stepped templates comprising two depths to simulate more complex boundary conditions in the spreading process.  These deeper regions lie $175$~\textmu m below the nominal template depth.  The sidewall angle of the templates is $54.7^\circ$ relative to the top surface of the wafer, corresponding to the orientation of the \textless 111\textgreater {} plane in \textless 100\textgreater {} oriented Si wafers, due to the anisotropic etching process employed.  The reader is referred to Section~\ref{app:WaferFab} for a full description of the template fabrication process.

Template depth is qualified post-fabrication by confocal microscopy (Keyence VK-X1050).  Figure~\ref{fig:Templates}c presents a typical measurement of the surface profile at the edge of a template edge (Profile A), and Fig.~\ref{fig:Templates}d shows a depth of 130.6 \textmu m has been achieved.  No conclusions are drawn here about the accuracy of the inclined surface due to the difficulty in optically measuring a highly reflective surface at a high angle; however, the average slope closely matches the expected value. Figure~\ref{fig:Templates}e shows the surface profile of a 1 mm square region in the bottom of the 131 \textmu m template; surface roughness parameters are $S_a = 0.033$ \textmu m and $S_z = 0.358$ \textmu m.

\subsection{Powders}

\begin{figure*}[htb]
\begin{center}
	{
	\includegraphics[trim = {1.1in 1.9in 1.1in 3.75in}, clip, scale=1, keepaspectratio=true]{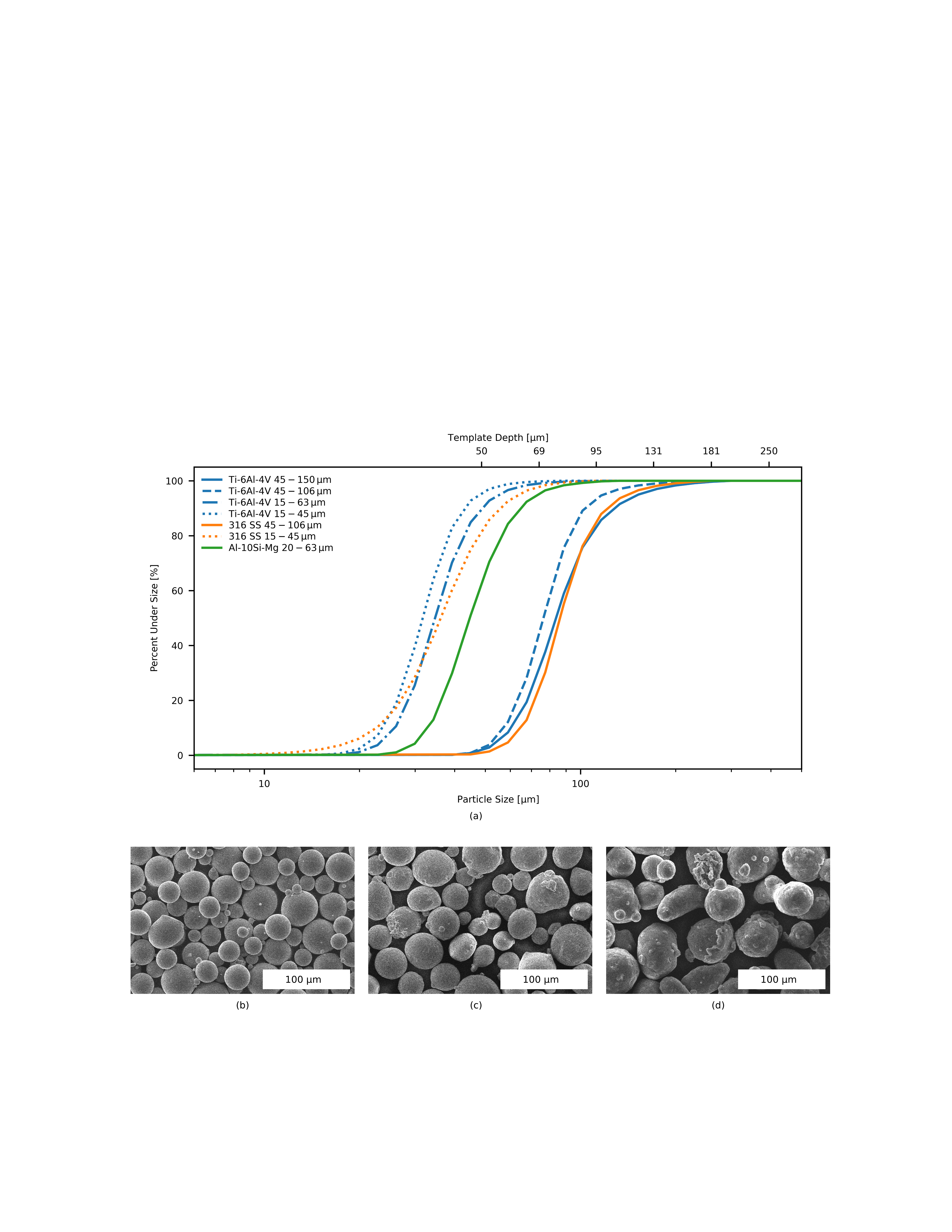}
	}
\end{center}
\vspace{-11pt}
\caption{Metal powders used herein.  (a) Size distributions measured via laser diffraction.  (b-d) SEM microscopy of 15-45~\textmu m Ti-6Al-4V, 15-45~\textmu m 316 SS, and 20-63~\textmu m Al-10Si-Mg, respectively.}
\label{fig:Powder}
\end{figure*}

Table~\ref{table:Method_Powders} lists the powders employed herein, selected to cover a range of common materials and size distributions in industrial PBF AM.  The nominal size parameters provided by each manufacturer are compared to distributions measured via laser diffraction (Horiba LA960), and the distributions are additionally plotted in Fig.~\ref{fig:Powder}a.  A selection of SEM images is shown in Figs.~\ref{fig:Powder}b through~\ref{fig:Powder}d; qualitatively, we observe that the Ti-6Al-4V and 316 SS powders are substantially spherical and the Al-10Si-Mg powder is comparatively oblate.  

\begin{figure}[htb]
\begin{center}
	{
	\includegraphics[trim = {.8in 2.5in 4.5in 3.1in}, clip, scale=1, keepaspectratio=true]{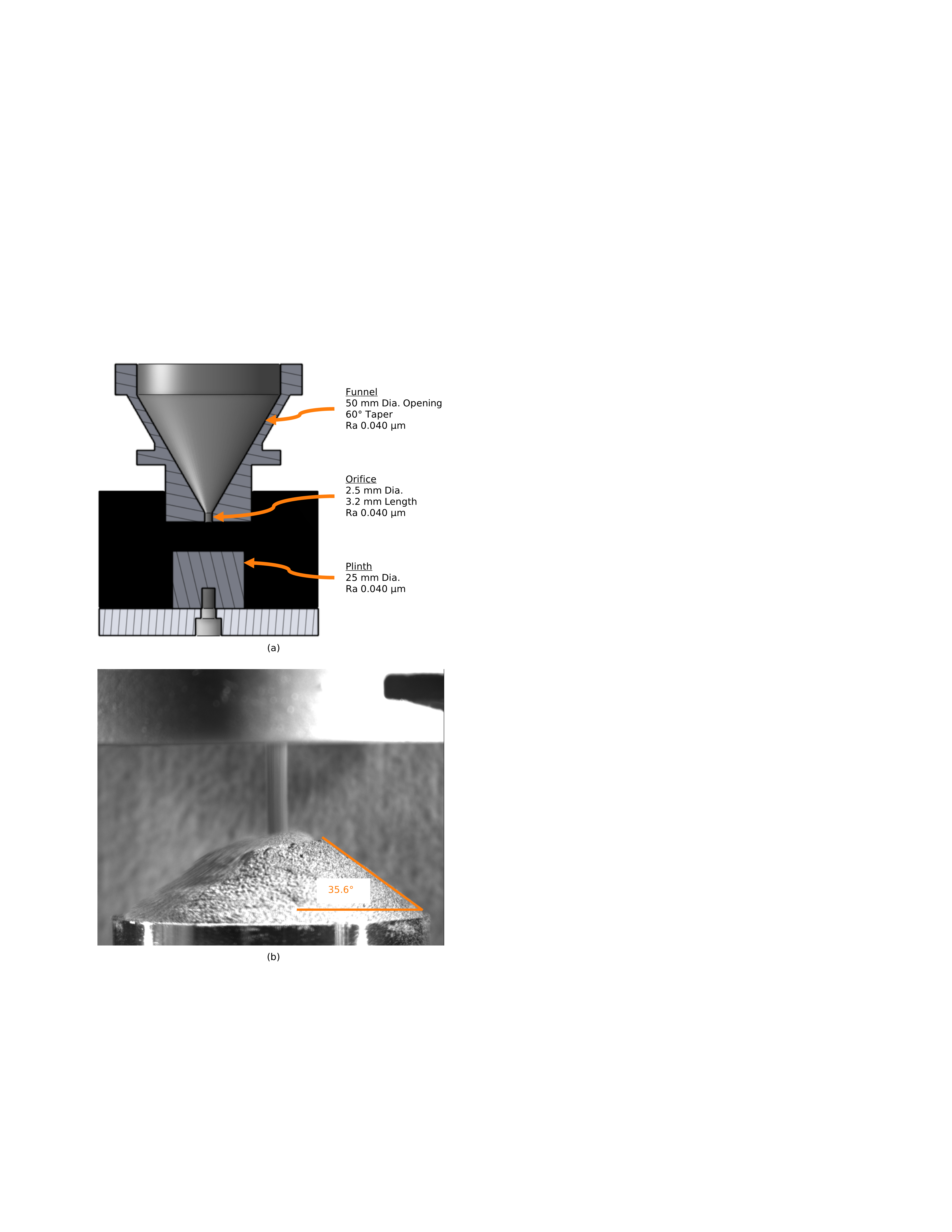}
	}
\end{center}
\caption{Angle of repose apparatus and measurement.  (a) Critical dimensions of the funnel and plinth.  (b) Still frame from a typical angle of repose measurement, illustrating angle measurement from a static region of the powder pile.}
\label{fig:AoR}
\end{figure}

Angle of repose measurements are also tabulated, performed in-house using a funnel inspired by a Hall Flowmeter Funnel (see~\cite{ASTM2013_ReposeAngle} and Fig.~\ref{fig:AoR}) to deposit powder on a $25$~mm diameter circular plinth.  Precise knowledge of the size distribution and the angle of repose are used in calibration of the simulations described below, and provide a notion of the significance of interparticle adhesive and frictional forces with respect to inertial (gravitational) forces.

\begin{table*}[htb]
\centering
\footnotesize
\caption{Tabulated powder properties.}
\label{table:Method_Powders}
\begin{tabular}{@{}*7c@{}}
  
  \toprule[1.5pt]
  \multicolumn{1}{c}{\head{Material}} &
  \multicolumn{1}{c}{\head{Nominal Size [\textmu m]}}&
  \multicolumn{1}{c}{\head{D10 [\textmu m]}} &
  \multicolumn{1}{c}{\head{D50 [\textmu m]}}&
  \multicolumn{1}{c}{\head{D90 [\textmu m]}}&
  \multicolumn{1}{c}{\head{Angle of Repose [$^\circ$]}}&
  \multicolumn{1}{c}{\head{Supplier}}\\
  
  \cmidrule{1-7}
 
    Ti-6Al-4V & 15-45 & 23.4 & 31.5 & 43.4 & 35.6 & AP\&C (GE) \\
    Ti-6Al-4V & 15-63 & 25.8 & 34.7 & 49.1 & 35.1 & AP\&C (GE) \\
    Ti-6Al-4V & 45-106 & 59.2 & 81.2 & 114.7 & 27.3 & AP\&C (GE) \\
    Ti-6Al-4V & 45-150 & 60.2 & 83.7 & 128.4 & 26.7 & AP\&C (GE) \\
    316 SS & 15-45 & 22.6 & 36.1 & 56.0 & 35.5 & Carpenter Technology \\
    316 SS & 45-106 & 62.4 & 83.9 & 117.4 & 34.0 & Carpenter Technology \\
    Al-10Si-Mg & 20-63 & 32.7 & 44.7 & 64.9 & 38.2 & IMR Metal Powders\\
 
  \bottomrule[1.5pt]
\end{tabular}
\end{table*}

\subsection{Powder Layer Fabrication}

Spreading of powder into the etched silicon wafer templates is performed manually using a 1/8 in.\ thick machinist parallel as a blade. The blade edge is rested on the wafer, several millimeters from the template to be filled, and a gross excess of powder is deposited immediately in front of the blade on the unetched surface.  The template is then filled by sweeping the blade across the template in one smooth motion, taking care that the blade is neither permitted to rise off the wafer nor rock as to change the right angle between the template and blade face.  Spreading speed is not measured, but is estimated to be $\approx 5$~mm/s.  This entire process is performed with the template attached to the X-ray specimen stage, as powder shifts if templates are moved.

Ambient humidity is known to affect powder spreading mechanics; while powder spreading is not performed under controlled humidity, extremes due to weather fluctuations are avoided. Temperature and humidity are measured when experiments are performed (Extech 42275).  Data logged during the experiments collected herein indicate an average temperature of $22.8^\circ$C (range $22.4^\circ$C-$23.2^\circ$) and an average relative humidity of 36.7\% (range 22.6\%-46.4\%).

\subsection{X-ray Imaging}
\label{sec:MethodsRTExposure}

X-ray imaging is performed with parameters selected in view of the material to be imaged (see Section~\ref{sec:MethodRTModel}) and the equipment operating range.  The source is set to a tube potential of $50$~kV and current of $200$~\textmu A, enabling use of the lowest selectable emission spot size ($5$~\textmu m). Detector exposure time is typically $1-10$~s; this is set empirically, such that maximum observed signal levels are approximately 80\% of the dynamic range of the detector, as to avoid saturation nonlinearities.  The number of counts collected in a single exposure (signal strength) is generally insufficient to resolve fine differences in material deposition, as discussed in section~\ref{sec:methods_RT_noise}.  Thus, frames are summed to achieve the required signal strength; the requisite number of frames is calculated by dividing the number of counts required by the mean number of counts recorded in a single image.

Determining the effective depth of a powder layer requires collection of two images (or sums thereof), namely a reference image $I_0$ of the empty template and a second image $I$ after spreading powder into the template.  Transmission $T$ is simply calculated as the ratio $I/I_0$.  This step cancels pixel-to-pixel sensitivity variation, as well as unknown photon and quantum efficiencies of the X-ray detector. 

Great care is taken in dark current (or signal) subtraction to ensure accurate results as, clearly, $I/I_0 \neq (I+\delta )/(I_0+\delta )$.  The considerable exposure times required make this problem especially significant, as the integrated dark current is proportional thereto and approaches 20\% of the detector's dynamic range.  Moreover, empirical observation shows that dark current magnitude is a strong function of temperature, which is poorly controlled due to ambient temperature fluctuations and nature of airflow within the X-ray cabinet.  Thus, the total number of frames required is divided into subsets, each representing approximately $5$~min of net exposure time, and dark frames are recorded before and after each subset to measure this offset.  Linear interpolation is then used to determine the dark current midway through each active exposure for subtraction.

\subsection{Radiation Transport Model}
\label{sec:MethodRTModel}

\begin{figure*}[ht]
\begin{center}
	{
	\includegraphics[trim = {0.8in 0.95in 0.8in 2.75in}, clip, scale=1, keepaspectratio=true]{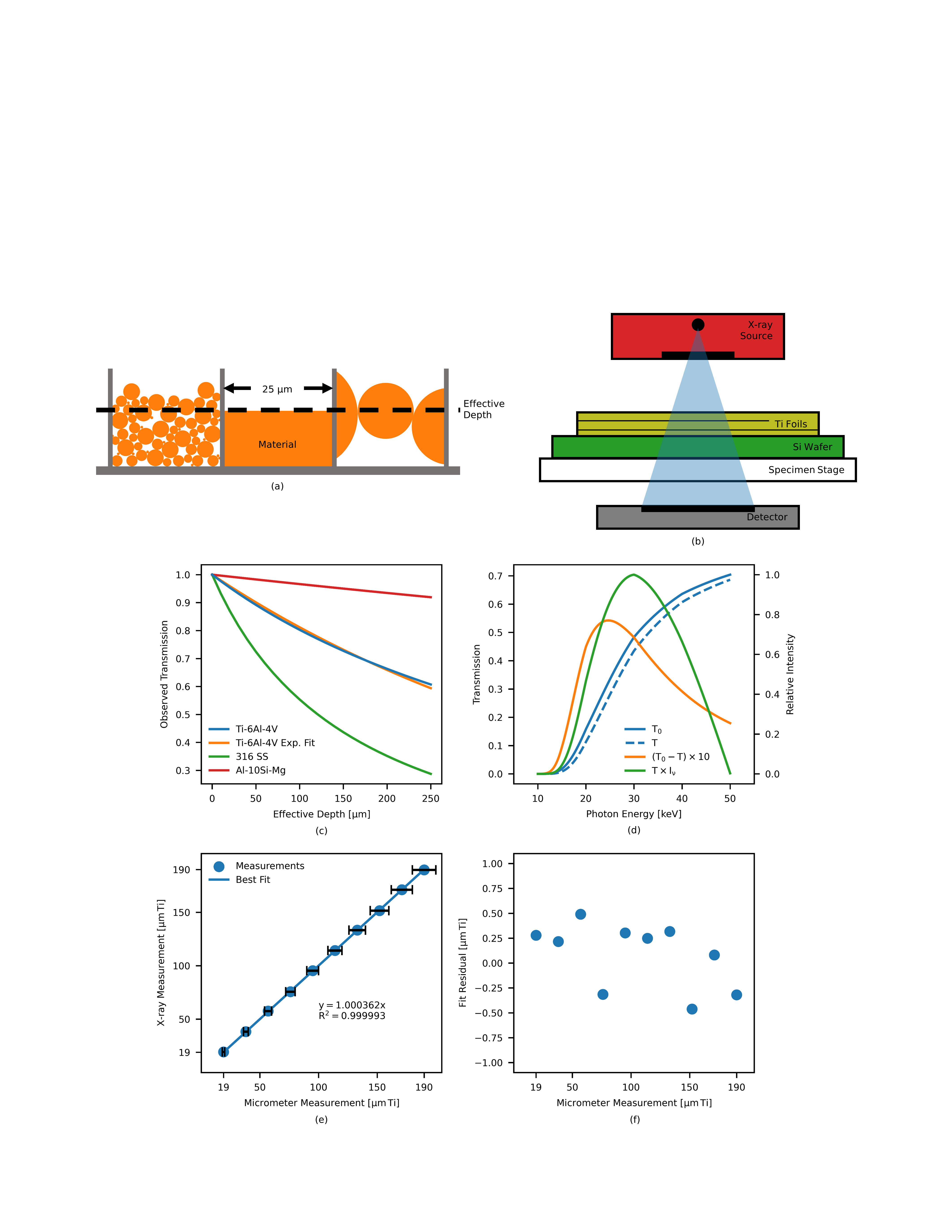}
	}
\end{center}
\caption{X-ray transmission measurement of powder layer effective depth. (a) Graphical definition of effective depth.  (b) Schematic illustration of the foil calibration experiment.  (c) Transmission vs. effective depth for studied powder materials.  (d)  Representative transmission spectra (blue), contrast (orange) generated by their difference, and spectrum of X-ray photons incident on the detector (green).  (e) Validation of X-ray measurements against calibration foils.  (f) Residuals from the line of best fit in (e).}
\label{fig:MethodX-ray}
\end{figure*}

A radiation transport (RT) model is used to interpret transmission measurements, as graphically represented in Fig.~\ref{fig:MethodX-ray}. This forward model calculates transmission as a function of powder layer effective depth (defined in Fig.~\ref{fig:MethodX-ray}a) for typical experimental parameters and materials of interest, as shown in Fig.~\ref{fig:MethodX-ray}c.  Relative to Ti-6Al-4V, for example, 316 SS is more dense and correspondingly shows greater X-ray stopping power.  Numerical inversion of these curves enables transmission measurements (Section~\ref{sec:MethodsRTExposure}) to be interpreted as effective depth of powder in the region sensed by each detector pixel.  The RT model explicitly considers the polychromatic nature of the X-ray source, wavelength-dependent attenuation coefficients of objects in the beam path, and physics of X-ray detection.  As evidenced by the functional difference between the Ti-6Al-4V curve and its exponential best fit, these effects are nonlinear and cannot simply be captured via simplistic application of Lambert's Law.

\subsubsection{X-ray Source Spectrum}

Our RT model computes X-ray source spectra based on the method of Birch and Marshall~\cite{Birch1979} (c.f.~\cite{Kramers1923, Soole1976, Tucker1991}), in which a physically-motivated framework is developed to describe spectra from commercial X-ray tubes.  Functionally, radiation is generated by bombarding a target with an electron beam.  A portion of the kinetic energy of the electrons is dissipated via the bremsstrahlung process, where radiation is generated as the electrons rapidly decelerate upon propagation into the target~\cite{Dyson1990}.  Birch and Marshall calculate the spectral intensity $I_\nu$ of bremsstrahlung radiation as a function of frequency $\nu$ per the integral
\begin{equation}
    I_\nu \propto \int_{V_0}^{V_\nu}\left(1+\dfrac{V}{m_0c^2}\right)Q\left(\dfrac{dV}{dx}\right)^{-1}exp\left(\dfrac{-\mu_\nu}{\rho C}(V_0^2-V_\nu^2)\cot{\theta}\right)dV
\end{equation}
where the limits run from the tube potential $V_0$ to $V_\nu$, or the potential that confers an electron kinetic energy equal to that of an X-ray photon with frequency $\nu$.  The first term under the integral gives a modest relativistic correction in view of the rest mass $m_0$ of an electron.  $Q$ is a differential energy intensity, as a function of the ratio of photon energy to to electron energy, where the empirically determined polynomial given in~\cite{Birch1979} has been used.  Electron stopping power, $dV/dx$, describes the rate of energy loss by electrons in propagating in the tungsten tube target; these data are interpolated as a function of electron energy from values from the NIST ESTAR database~(\cite{NIST1993}).  The final exponential term describes self-attenuation of X-rays as they propagate from their sub-surface point of genesis within tungsten target.  The point of X-ray genesis is calculated from the expected electron penetration depth (d), per the Thomson-Whiddington relation~\cite{Whiddington1914}
\begin{equation}
d=(V_0^2-V_\nu^2)/\rho C
\end{equation}
where $\rho$ is the target mass density and $C$ is a constant (interpolated from those provided by Birch and Marshall).  Target geometry, specifically the target angle $\theta$, increases the path length for X-rays to escape the target by a factor of $\cot{\theta}$.  This angle is estimated here as $\approx 23^\circ$ from observation of the heel effect.  Finally, $\mu_\nu$ is the attenuation coefficient of the target material at the frequency of interest.

Notably, we have elected not to consider generation of the characteristic (K-shell) X-ray emissions of the target, despite being incorporated into the original Birch and Marshall model.  Electron kinetic energy at the selected tube potential (50~keV), as motivated below, lies well below the minimum K-shell binding energy of tungsten ($69.5$~keV~\cite{Bearden1967}). Therefore, these selective emission peaks cannot be excited~\cite{Dyson1990} and no fidelity is lost in their omission from our analysis.

\subsubsection{Transmission}

The second component of the RT model calculates transmission spectra according to the presence of objects in the beam path, including the powder layer, between the source target and detector scintillator. For each object, transmission is determined in accordance with Lambert's Law
\begin{equation}
T = e^{-\mu l},
\label{eq:Lambert}
\end{equation}
where $T$ is transmission as a function of the attenuation coefficient, $\mu$, and path length (object thickness) $l$.  Critically, $\mu$ depends both on the material and on photon energy \cite{Compton1940}; thus these calculations are performed as a function of wavelength using attenuation coefficients compiled by NIST (\cite{Saloman1988X-ray92}) for elements and some common materials.  Finally, the net transmission spectrum is determined simply by multiplying the spectra calculated for each object.

Attenuation coefficients for materials (e.g., powder alloys) that are not present in the database are calculated from elemental data via the mixing rule
\begin{equation}
    \dfrac{\mu}{\rho_m} = \sum_i w_i\left(\dfrac{\mu}{\rho}\right)_i
\end{equation}
where $w_i$ is the mass-fraction of the $i^{th}$ element comprising the material~\cite{Dyson1990, Saloman1988X-ray92}.  Accurate knowledge of powder composition is critical to ensuring accuracy in converting X-ray data to layer thickness.  It is insufficient to simply use the nominal composition; rather, the manufacturer-reported data for the specific powder lots including the presence of minor alloy elements must be used.  For example, even the trace amount ($\leq 0.5\%$) of copper typically present in 316 SS greatly impacts X-ray absorption.

Figure~\ref{fig:MethodX-ray}d shows two exemplary transmission spectra calculated with this technique.  The first curve, denoted $T_0$, represents net transmission through the beryllium source window, aluminum source guard, empty silicon template, polyoxymethylene sample stage, and carbon fiber composite detector cover.  An identical calculation further incorporating a Ti-6Al-4V powder layer with $50$~\textmu m effective thickness results in the second curve, $T$.  Multiplying the transmission spectrum with the source emission spectrum, as calculated above, generates an intensity spectrum of X-ray photons incident upon the detector scintillator ($T \times I_\nu$, green).

As an aside, subtracting the transmission curves allows one to understand contrast as a function of photon energy.  The scaled difference of the transmission curves in Fig.~\ref{fig:MethodX-ray}d, $\left(T_0-T\right)\times10$, shows that contrast peaks in the $20-25$~keV range for typical experiments.  Intuitively, soft (low-energy) X-rays generate little contrast, or convey little information about the layer thickness, because they are nearly completely absorbed by the materials between the source and detector.  Likewise, high energy photons also generate little contrast due to the low likelihood that they interact with (are specifically absorbed by) the powder layer.

\subsubsection{Detector}

The last components of the model account for the quantification of X-rays by the detector.  Detection occurs by first converting X-ray photons into visible light using a microcolumnar cesium iodide scintillator.  The light yield of this process is approximately $52$ visible photons per keV of X-ray photon energy~\cite{Holl1988}.  Second, there is a gain applied between the number of photons collected and the number of counts reported by the detector; this value is $2.00$~counts/photon as determined from noise statistics.

\subsubsection{Integration Time and Noise Statistics}
\label{sec:methods_RT_noise}

Integration time, or proportionally the aggregate signal reported by the detector in counts, is set as to provide a sufficient signal-to-noise ratio to resolve a 1~\textmu m difference in the material of interest.  This is calculated using the formula for error propagation,
\begin{equation}
    \sigma_T = T\sqrt{\left(\dfrac{\sigma_{I_0}}{I_0}\right)^2+\left(\dfrac{\sigma_I}{I}\right)^2}
\end{equation}
where $\sigma$ represents the standard deviation of the quantity indicated by the subscript.  Observation of measurement noise indicates that the signal-to-noise ratio is strongly limited by (Poisson distributed) shot noise, allowing the substitutions $\sigma_{I_0} = \sqrt{I_0}$ and $\sigma_I = \sqrt{I}$.  Further taking $I \approx I_0$, or that small changes in material thickness cause small changes in $I$, allows reduction and rearrangement of the formula to
\begin{equation}
    I \approx \dfrac{2}{\sigma_T^2}.
\end{equation}
Thus, if a 1~\textmu m increase in titanium thickness changes transmission by $2.6\times 10^{-3}$, $I$ must be $\geq 3.0\times10^5$ counts to resolve the difference.

\subsubsection{Validation}
\label{sec:methods_RT_cal}

The RT model is validated through X-ray attenuation measurement of stacks of titanium foils (Goodfellow TI000251), as schematically illustrated in Fig.~\ref{fig:MethodX-ray}b, using exposure parameters determined via the methods above.  According to the manufacturer specifications, each foil is nominally $20 \pm 3$ \textmu m thick; our measurement of the thickness using a micrometer (Mitutoyo 293-831-30) gives $19 \pm 1$ \textmu m.  Examining the thickness of a single foil over its area, the X-ray technique estimates an average thickness of $19.3\pm1$~\textmu m that compares favorably with the micrometer measurement and expected uncertainty due to shot noise.  Figure~\ref{fig:MethodX-ray}e plots the full range of results for stacks of one to ten foils, including a line of best fit that indicates excellent linearity between X-ray and micrometer measurements.  Residuals arising from an unknown combination of variation in foil thickness and uncaptured physics are plotted in Fig.~\ref{fig:MethodX-ray}f.  All residuals are small as compared to the 1~\textmu m shot-noise measurement uncertainty, thus model precision is taken to be sufficient. Therefore, we consider the RT model and X-ray microscope's performance to be validated to 1~\textmu m and capable of accurate measurements of effective material thickness in powder layers.

\subsection{DEM Modeling of Powder Spreading}
\label{sec:MethodDEM}
We compare our physical experiments to a Discrete Element Method (DEM) computational model of powder spreading recently developed by Meier et al.~\cite{Meier2019CriticalManufacturing,Meier2019ModelingSimulations}; DEM was originally developed by Cundall and Strack \cite{Cundall1979}. The DEM powder model considers each powder particle individually including solving their equations of motion. The particles are assumed to be spherical, which is a suitable approximation for the considered gas-atomized powders (see Figs.~\ref{fig:Powder}b-d). Our DEM implementation considers particle-to-particle and particle-to-wall (e.g., between particles and the solid substrate or the recoating blade) interactions, including frictional contact, rolling resistance, and cohesive forces. As discussed in \cite{Meier2019ModelingSimulations}, the \textit{effective} surface energy underlying the model for adhesive interaction forces between powder particles can vary by several orders of magnitude for one given powder material, due to the roughness and potential chemical contamination of powder particle surfaces. Therefore, this crucial model parameter cannot be taken from standard databases, but has to be calibrated for the powder material at hand. In this study, the aforementioned static angle of repose measurements are used to calibrate the effective surface energy for each powder, as described in \cite{Meier2019ModelingSimulations}. In addition to an accurate representation of the particle interaction forces, a realistic representation of the particle size distribution is required. For this purpose a log-normal distribution is assumed and fitted to the data in Fig.~\ref{fig:Powder}a.

\begin{figure}[ht]
\begin{center}
	{
	\includegraphics[trim = {2.55in 3.2in 2.55in 3.2in}, clip, scale=1, keepaspectratio=true]{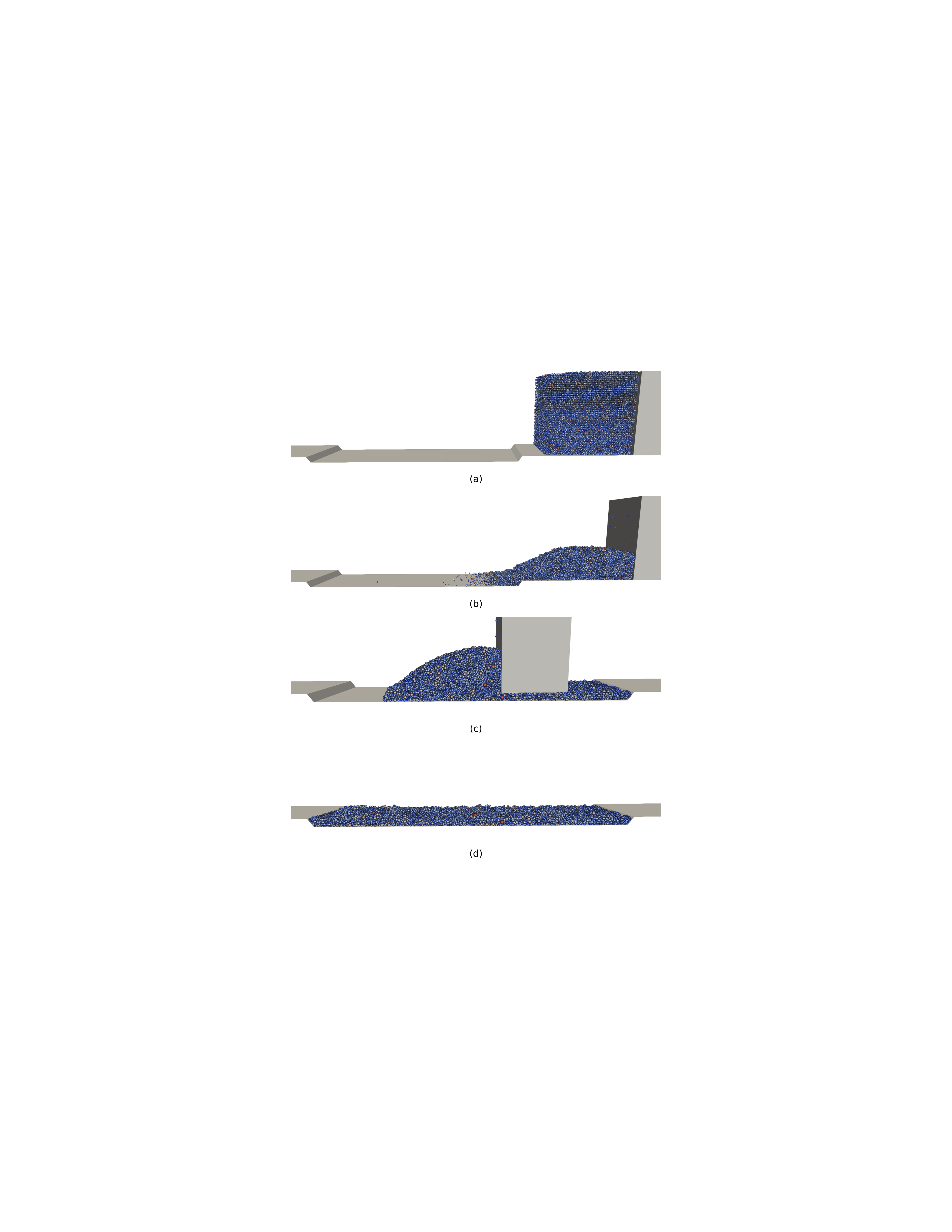}
	}
\end{center}
\caption{Simulation of spreading a single powder layer (see Supplementary Videos~1 and~2).  (a) Initial configuration (t=0~ms).  (b) Configuration after settling of the particles (t=40~ms).  (c) Configuration during spreading (t=1000~ms).  (d) Final configuration (t=1900~ms).}
\label{fig:SimAConfigurations}
\end{figure}

The process of spreading the powder into the silicon templates is simulated with this computational model. To limit the computational time, a representative domain is chosen out of the physical template area of $20\times 20$~mm. For the fine powders ($D_{50}<50$~\textmu m) the geometry is chosen as $5\times 1$~mm and for the coarser powders ($D_{50}>50$~\textmu m) as $20\times 2$~mm, where in both cases the longer (5~mm, respectively 20~mm) edges of the domain are in spreading direction.  Periodic boundary conditions are applied on the longer edges of the powder bed, such that particles leaving the domain on one side enter the domain on the other side. We checked the sensitivity of the choice of the $5\times 1$~mm domain by doubling its size and found no significant difference in the results. Finally, the standard deviation of the mean effective depths of five simulations with independently generated particle populations was less than 1~\textmu m for each layer thickness.  

For each DEM simulation (e.g., as shown in Supplementary Video~1), in the initial configuration, particles are arranged on a Cartesian grid with an additional random offset (Fig.~\ref{fig:SimAConfigurations}a). At the start of the simulation the particles are allowed to settle due to gravity (Fig.~\ref{fig:SimAConfigurations}b), to create a realistic powder pile in front of the blade as basis for the spreading. After the particles have settled, a blade sweeps the powder pile across the template at 5~mm/s (see Fig.~\ref{fig:SimAConfigurations}c), resulting in the final layer depicted in Fig.~\ref{fig:SimAConfigurations}d. A blade thickness of 1~mm was selected in view of the reduced length of the representative domain.  Two multi-layer simulation cases are considered, differing in the application of the first layer. For the first case, the particles of the first layer are placed directly into the template.  Thereto, the particles are initially placed in a grid with random offset and then settled by gravity. After the particles settle into the first layer, the excess particles are removed with a blade that moves perpendicularly to the spreading direction of the second layer.  The second case spreads powder into the lower level of the template, analogous to the spreading of the single layer simulations, to create the first layer. For both case the second layer is then spread with blade motion parallel to the long dimension of the template.

Numerous powder layer metrics may be extracted from post-processing of the DEM simulation results, such as packing fraction and surface profile~\cite{Meier2019ModelingSimulations,Meier2019CriticalManufacturing}.  Comparison to X-ray results is possible by evaluating the virtual domain in the same fashion as in the experiment. Therefore, rays are defined on a regular grid with resolution 5~\textmu m, simulating the X-rays. The effective depth consists of the accumulated length of the intersections of the rays with the particles, and the effective depth measured by each ray is averaged inside a 25~X~25~\textmu m$^2$ domain, such that the same resolution as in the experiment is achieved. Thus, these data are directly analogous to the X-ray data and are therefore analyzed identically. Additionally, simulation results provide insights, such as particle motion during spreading, that are not resolved in the physical experiment.

\section{Results}
\label{sec:results}

\begin{figure*}[h!tb]
\begin{center}
	{
	\includegraphics[trim = {2in 3.25in 1.55in 3.3in}, clip, scale=1, keepaspectratio=true]{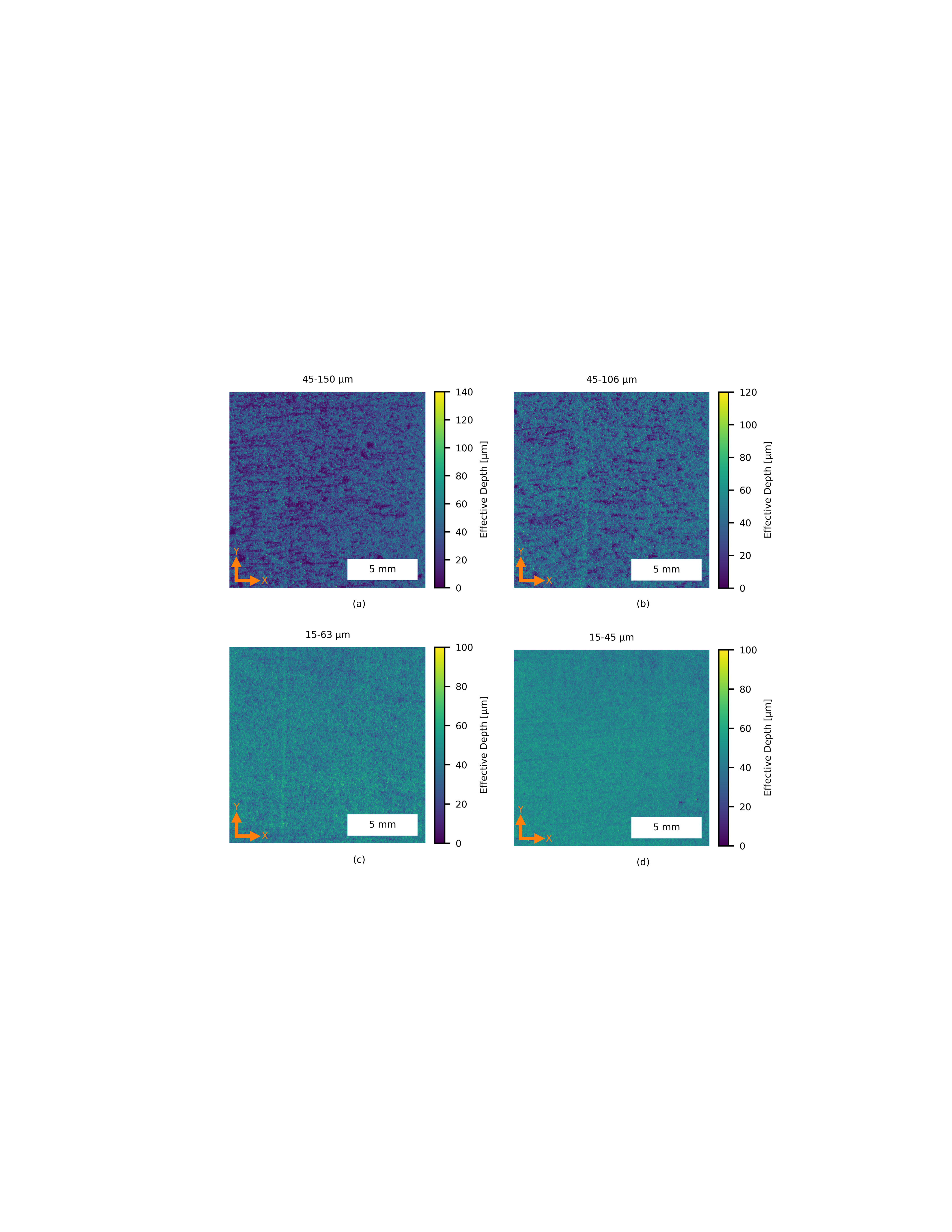}
	}
\end{center}
\caption{Typical layers of a range of Ti-6Al-4V powder sizes spread in the -X direction (from right to left) into a 131~\textmu m deep template, imaged via X-ray microscopy.}
\label{fig:ResultsLayers}
\end{figure*}

Using the materials and methods described above, we create powder layers that span a range of materials and layer thicknesses, and with differing boundary conditions. Spreading is first studied in uniform-depth templates, selectively considering the central $14\times 14$~mm region, and therefore removing spatial variation due to perturbed powder flow at template boundaries.  Representative layers of Ti-6Al-4V spread into a 131~\textmu m template are shown in Fig~\ref{fig:ResultsLayers}; qualitatively, one may observe that the finer (15-45~\textmu m and 15-63~\textmu m) Ti-6Al-4V powders result in the densest coverage (see corresponding simulation in Supplementary Videos~1 and~2).  Conversely, the 45-106~\textmu m Ti-6Al-4V powder shows structured defects, primarily streaks caused by large powder particles that cannot easily pass between the blade and template (see corresponding simulation in Supplementary Videos~3 and~4).  This trend continues in the 45-150~\textmu m Ti-6Al-4V powder, where streaks and bare regions are plainly evident and only the fine particles of the powder distribution are deposited into the template.  

In the following sections, these observations are quantified through determination of packing fraction, statistical distribution of effective depth, and defect sizing in the frequency domain.  Finally, disturbance of powder flow about abrupt changes in geometry are examined using stepped templates to gain preliminary insights into the uniformity of multi-layer spreading.

\subsection{Packing Fraction}

\begin{figure*}[ht]
\begin{center}
	{
	\includegraphics[trim = {0.65in 4.75in 1.15in 2.95in}, clip, scale=1, keepaspectratio=true]{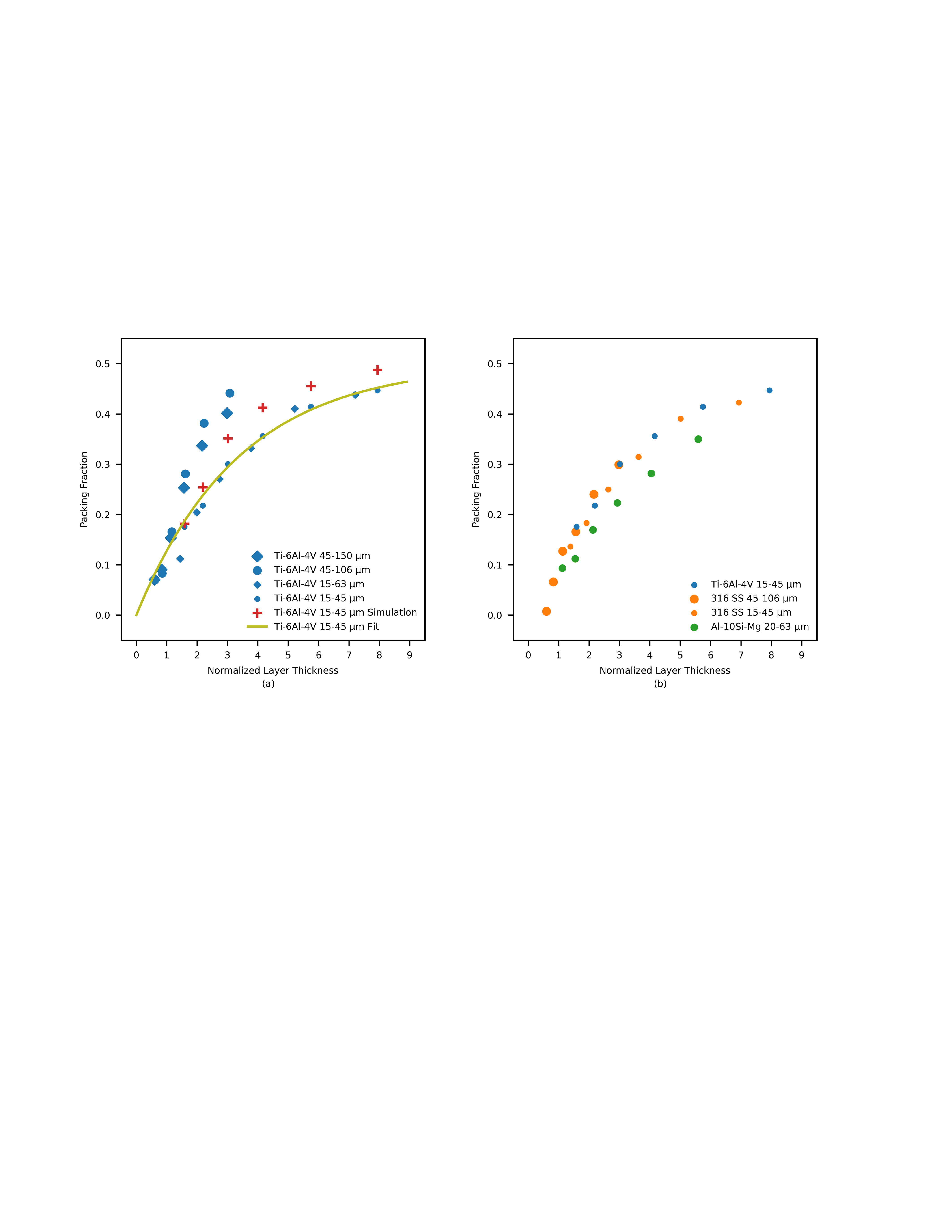}
	}
\end{center}
\caption{Packing fraction versus normalized layer thickness.  Note, 15-45~\textmu m Ti-6Al-4V powder data are reproduced in (a) and (b) for comparison.}
\label{fig:ResultsPackingFraction}
\end{figure*}

Analysis begins by calculating the mean packing fraction of the powder layers, defined as the average effective depth divided by the template depth.  Figure~\ref{fig:ResultsPackingFraction} plots these data against normalized layer thickness, calculated as the template depth divided by the powder $D_{50}$, thereby providing a relative metric of how deep the template is as compared to the powder spread into it.  Relatively similar spreading behavior is found for the 15-45~\textmu m and 15-63~\textmu m Ti-6Al-4V powders as well as the 15-45~\textmu m and 45-106~\textmu m 316 SS powders.  We rationalize this by the narrow range of angle of repose ($34^\circ$ to $35.6^\circ$) for these powders, indicating similar flowability and, hence, spreadability.  At high normalized layer thickness, packing fraction approaches an asymptotic value; for example, fitting an exponential (functionally $\alpha\left( 1-e^{-\beta}\right)$, gold in Fig.~\ref{fig:ResultsPackingFraction}a) curve to the 15-45~\textmu m Ti-6Al-4V data implies saturation at a packing fraction of $0.50$.  This value compares favorably to a packing fraction of $0.56$ calculated from the manufacturers reported tap density, which places an upper bound on the value expected here.  Figure~\ref{fig:ResultsPackingFraction}a additionally shows simulation results corresponding to the 15-45~\textmu m Ti-6Al-4V data.  DEM simulations reproduce this relationship, albeit with a slight shift in numerical values, namely saturation at a packing fraction of 0.55 that is slightly below the (separately) simulated tap packing fraction of 0.58.

However, the combination of powder material and size distribution affects the characteristics of layers via changes in the relative magnitude of gravitational and van der Waals forces.  In Ti-6Al-4V, larger powder sizes feature considerably lower angle of repose as compared to the fine size distributions, indicating that increased particle mass works to overcome adhesive forces and leads to settling into dense layers as a result. In the opposite direction, the Al-10Si-Mg powder is observed to be the least flowable (AoR of $38.2^\circ$) and creates the most sparse layers.  This indicates strong cohesive forces as compared to gravitational forces, explained by the low mass density of aluminum, and further implies clumping of powder particles during spreading.

At low normalized layer thickness the measured packing fractions of all powders studied converge, as driven by two key considerations.  First, only the fine particles of the distributions are deposited, e.g., all layers resemble Fig.~\ref{fig:ResultsLayers}a.  Second, thin layers lead to comparatively high-shear conditions that break up clusters of powder particles.

Critically, typical PBF AM spreading conditions span the full range of normalized layer thickness presented here. In a typical LPBF process employing a solidified layer thickness of 50 \textmu m and 15-45\textmu m Ti-6Al-4V powder ($D_{50}=31.5$~\textmu m), the first layer, as well as powder spread over unfused material in subsequent layers experience a normalized layer thickness of $50/31.5 \approx 1.6$.  As the build continues, the actual thickness of spread powder above previously fused material approaches the nominal layer thickness divided by the packing fraction, as described by Mindt and coworkers~\cite{Mindt2016}.  Therefore, a powder layer of $50/0.35 = 143$~\textmu m is necessary to form a solidified layer of 50~\textmu m thickness, using the packing fraction data presented in Fig.~\ref{fig:ResultsPackingFraction}, and the corresponding normalized layer thickness is $\approx 4.5$. 

\subsection{Effective Depth Distribution}

\begin{figure}[htb]
\begin{center}
	{
	\includegraphics[trim = {2.55in 2.75in 2.75in 3in}, clip, scale=1, keepaspectratio=true]{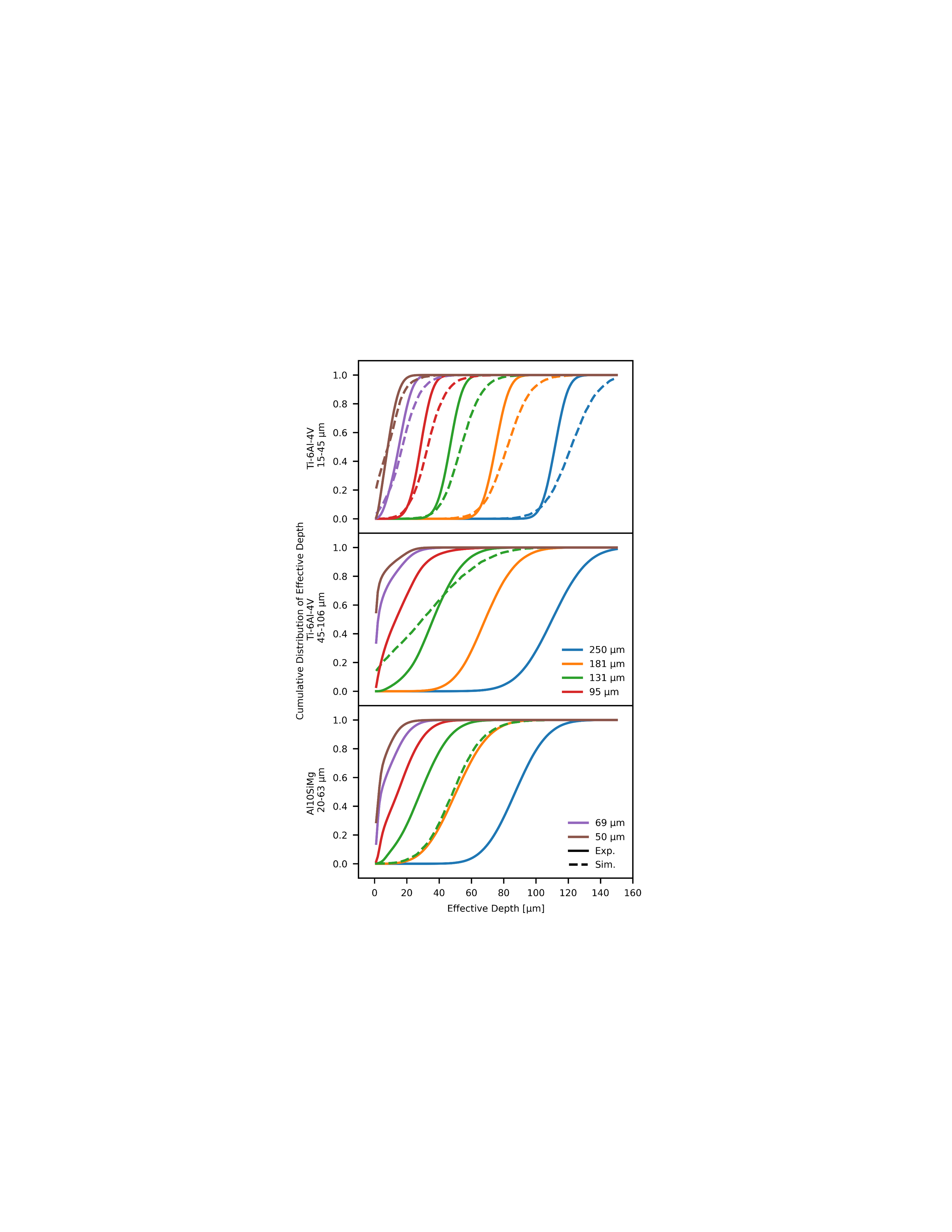}
	}
\end{center}
\caption{Cumulative distributions of effective depth as a function of layer depth and powder.}
\label{fig:ResultsHistograms}
\end{figure}

Figure~\ref{fig:ResultsHistograms} provides a more detailed statistical view of powder deposition through depiction of cumulative effective depth distributions. The top panel centers on 15-45 \textmu m Ti-6Al-4V, showing an effectively Gaussian distribution of effective depth.  Experiments at low nominal layer thicknesses, namely 50 and 69 \textmu m, show narrower distributions, whereas the distribution width is larger and independent of layer thickness for the thicker layers.  DEM simulation results show close correspondence to mean effective depth for thin layers, but modestly differ in over-predicting effective depth and the distribution width thereof for thick ($\geq 95$~\textmu m) layer conditions.  We hypothesize that the wider blade used in the experiments as compared to the simulated blade results in more consistent settling of the powder through more blade-particle interactions, evidenced in part by particle motion under the recoating blade in Supplementary Videos~1 through~4, and paralleling experimental results by Beitz et al.\ in PA12 nylon~\cite{Beitz2019}.

The middle panel of Figure~\ref{fig:ResultsHistograms} illustrates cumulative distributions for layers of 45-106 \textmu m Ti-6Al-4V, where the probability of bare regions is significant for the shallower template conditions. The widths of these distributions are considerably larger as compared to the fine Ti-6Al-4V powder, indicating more layer variability, and the means shift lower.  Moreover, these changes are reproduced in the corresponding DEM simulations.

The bottom family of curves in Figure~\ref{fig:ResultsHistograms} is for Al-10Si-Mg powder layers.  Expectations based on particle size would suggest spreading behavior bounded by that exhibited by the Ti-6Al-4V powders discussed above.  However, the effective depth distributions more closely mimic the larger Ti-6Al-4V powder in distribution width and with increased probability of bare regions, suggesting that this more cohesive powder resists flow into a dense and uniform configuration.  Simulations agree, as evidenced by an effective depth distribution that is approximately twice as wide as the 15-45~\textmu m Ti-6Al-4V powder.

\begin{figure*}[htb]
\begin{center}
	{
	\includegraphics[trim = {1in 2.75in 1.3in 4.75in}, clip, scale=1, keepaspectratio=true]{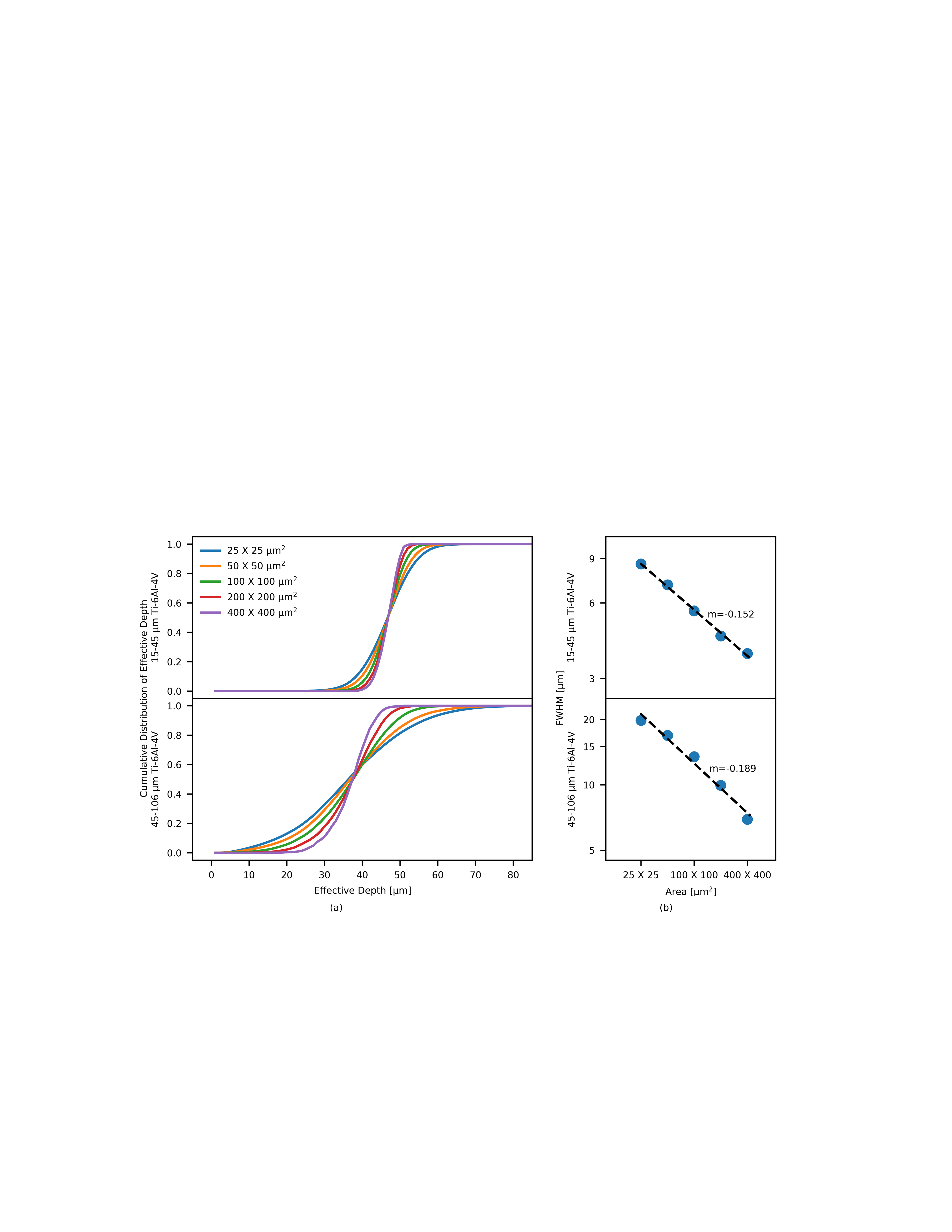}
	}
\end{center}
\caption{(a) Cumulative distributions of effective depth as a function of layer depth and area considered. (b) Corresponding full width at half maximum.}
\label{fig:AreaHistograms}
\end{figure*}

Finally, to provide an alternate representation of these data, we present averages of effective depth over progressively larger areas in Fig.~\ref{fig:AreaHistograms}, obtained by rebinning the effective depth measurements (i.e., Figs.~\ref{fig:ResultsLayers}b and~\ref{fig:ResultsLayers}d)
over a range of coarser spatial resolutions.  The rebinned images may be interpreted via cumulative distributions, as above, and are plotted in Fig.~\ref{fig:AreaHistograms}a.  With increasing bin size, the width of the effective depth distributions falls, here for 15-45~\textmu m and 45-106~\textmu m Ti-6Al-4V powder in a 131~\textmu m template. This representation is a useful comparison to the relevant length scales of processing in PBF AM. As such, selection of processing methods that act over larger areas, e.g., BJ with large droplet size or LPBF with a large spot size, may reduce but not entirely mitigate the impact of effective depth fluctuations on process stability and local component defects.

Figure~\ref{fig:AreaHistograms}b plots the full width at half maximum (FWHM) of the aforementioned distributions versus the averaged area.  We observe the FWHMs corresponding to the fine (15-45~\textmu m) powder layer fall from 9~\textmu m to roughly 4~\textmu m, in a linear fashion with a slope of -0.152 on logarithmic axes, as the area considered increases from 25$\times$25~\textmu m$^2$ to 400$\times$400~\textmu m$^2$.  The distributions corresponding to the large (45-106~\textmu m)  powder show a similar trend, decreasing from a value of approximately 20~\textmu m when evaluating effective depth over a 25$\times$25~\textmu m area with a mean slope of -0.189 for the region sizes considered.  These observations may be contrasted to consideration of effective depth as an independent Gaussian random variable, in which case the distribution width would have a -0.5 slope with increasing area (or proportionally increasing number of pixels or measurements averaged).  We conclude, based upon the markedly lower slope in the experimental data, that spatially-structured defects in powder layers (e.g., streaks) are not only present, but are significant in magnitude as compared to unstructured variations (e.g., stochastic sphere packing) at these length scales.

\subsection{Defect Morphology and Powder Layer Variance}

\begin{figure}[htb]
\begin{center}
	{
	\includegraphics[trim = {2.4in 1.75in 3in 1.25in}, clip, scale=1, keepaspectratio=true]{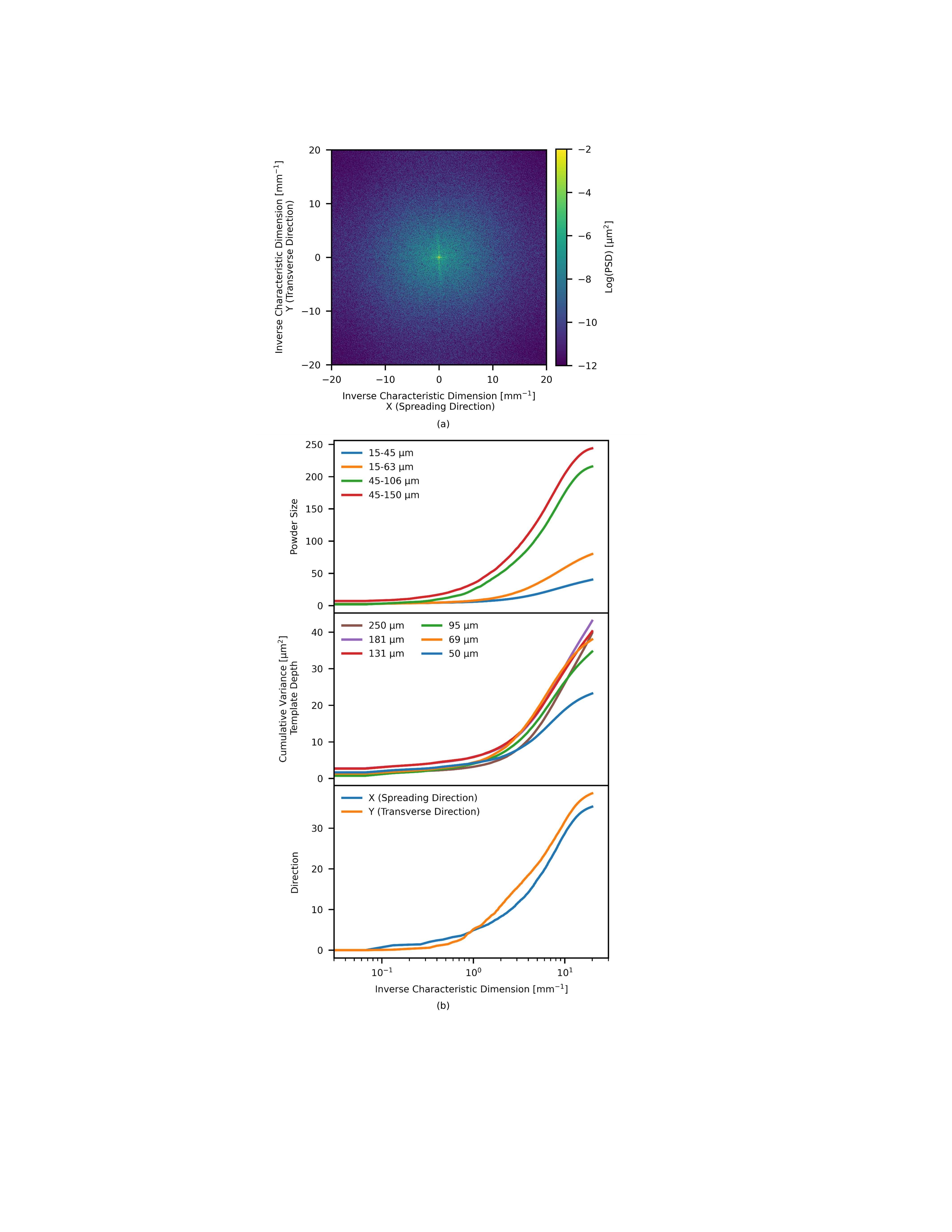}
	}
\end{center}
\caption{PSD analysis of powder layer variance.  (a) Typical PSD corresponding to the layer depicted in Fig.~\ref{fig:ResultsLayers}d.  (b) Comparison of Ti-6Al-4V layer variance versus defect size for (top) a range of powder sizes in a 131~\textmu m template, (center) 15-45~\textmu m powder in a range of template depths, and (bottom) ascription of variance to defect orientation of 45-106~\textmu m powder in a 131~\textmu m template.}
\label{fig:PSD}
\end{figure}

With the presence of spatially correlated fluctuations in effective depth established above, we turn to frequency-domain methods to quantify powder layer defect size and severity.  Herein, we first apply a discrete Fourier transform (DFT) to the mean-subtracted effective depth measurement.  The DFT is then squared to yield a 2-dimensional power spectral density (PSD, $S$); this is shown in Fig.~\ref{fig:PSD}a corresponding to the 131~\textmu m layer of 15-45~\textmu m Ti-6Al-4V powder depicted earlier in Fig.~\ref{fig:ResultsLayers}d.  

The DFT maps variance in powder layer depth to defect orientation and characteristic size.  For example, variance from large defects with low inverse characteristic dimension (analogous to low frequencies in the DFT of a time-domain signal) are mapped to pixels near the origin of the plot.  Conversely, high spatial frequency components arising from small defects are mapped closer to the perimeter.  In the context of LPBF, for example, these data inform powder-driven meltpool disturbance as a function of laser focus size.  Alternatively, the degree of large and intermediate scale powder layer variance that may be rejected via closed-loop control is also readily determined, with the intersection of controller bandwidth and laser scan speed defining a characteristic length scale.

To assist in interpreting the power spectral data, Fig.~\ref{fig:PSD}b plots cumulative variance (denoted here as power, $P$) versus characteristic dimension, calculated per the discrete equivalent of
\begin{equation}
    P(r) = \int_0^{2\pi} \int_0^r S(r',\theta') dr'd\theta'
    \label{eq:PSD}
\end{equation}
where polar coordinates are chosen to simplify the integral.  The top panel of this figure illustrates the influence of powder size on variance of effective depth in 131~\textmu m Ti-6Al-4V layers.  Larger powders have greater total variation in effective depth and the geometric extent of the defects are larger than those present in finer powders.  Specifically, we see that half of the observed variance of the 45-150~\textmu m and 45-106~\textmu m powder arises at inverse defect size of 4.6 and 5.1~mm$^{-1}$, respectively (220~\textmu m and 190~\textmu m defects).  The corresponding point for the finer powders lies at $\approx 6$~mm$^{-1}$, or 170 \textmu m.  We hypothesize that this non-proportional change in defect versus powder size arises both from higher adhesive forces in the smaller powders, causing clustering that persists during spreading, and the qualitative observation that streaks are orders of magnitude longer than typical particle diameters.  Finally, low variance is observed across all layers when considering inverse characteristic dimensions below $\approx 0.2$~mm$^{-1}$ (or 5~mm defect dimension).  Thus, while the layers of large powders in Fig.~\ref{fig:ResultsLayers} may show sparse deposition, these layers are as uniform as layers of finer powder over long ($\geq5$~mm) distances.

In the middle panel, layers of 15-45~\textmu m Ti-6Al-4V are studied in a variety of template depths.  A rough trend towards increased total variance with increasing layer thickness is seen, especially as compared to the thinnest layer (50~\textmu m), as particles can settle into more complex configurations in thicker layers.  Based upon the similarity of the curves from the thicker layers, characteristics (severity and size) of defects appear to be independent of thickness for sufficiently deep layers.

The final panel of Fig.~\ref{fig:PSD}b further examines a 131~\textmu m thick layer of 45-106~\textmu m powder, selected from the exemplary layers because the particles (and streaks resulting therefrom) are large as compared to the imaging resolution and are therefore most clearly resolved.  The two curves in the figure are calculated using an adapted version of Eq.~\ref{eq:PSD},
\begin{equation}
    P(r) = 2\int_{\theta_1}^{\theta_2} \int_0^r S(r',\theta') dr'd\theta'
    \label{eq:PSD2}
\end{equation}
where the \straighttheta{} integral limits restrict the integral to a sector and the factor of two necessarily accounts for the opposite and equivalent range of directions (the reader is reminded that both the Fourier transform and PSD of a real-valued function is symmetric and both sides must be integrated).  Using this relation, the PSD is summed in the sector spanning $0\pm 15^\circ$, aligned to the spreading direction, as well as the transverse sector spanning $90\pm 15^\circ$.  At an intermediate inverse characteristic dimension, equivalently between 1 and 10 millimeters, we see that more variance is encountered moving along the spreading direction than orthogonal to it.  Conversely, more variance occurs in the transverse direction at high inverse characteristic dimension, specifically at length scales approaching those of the largest powder particles in the size distribution.  This matches with a qualitative reading of Fig.~\ref{fig:ResultsLayers}b, clearly featuring numerous horizontal streaks of approximately this orientation and size.

\subsection{Multi-Layer Spreading}

\begin{figure*}[htb]
\begin{center}
	{
	\includegraphics[trim = {0.85in 1.3in .75in 1.40in}, clip, scale=1, keepaspectratio=true]{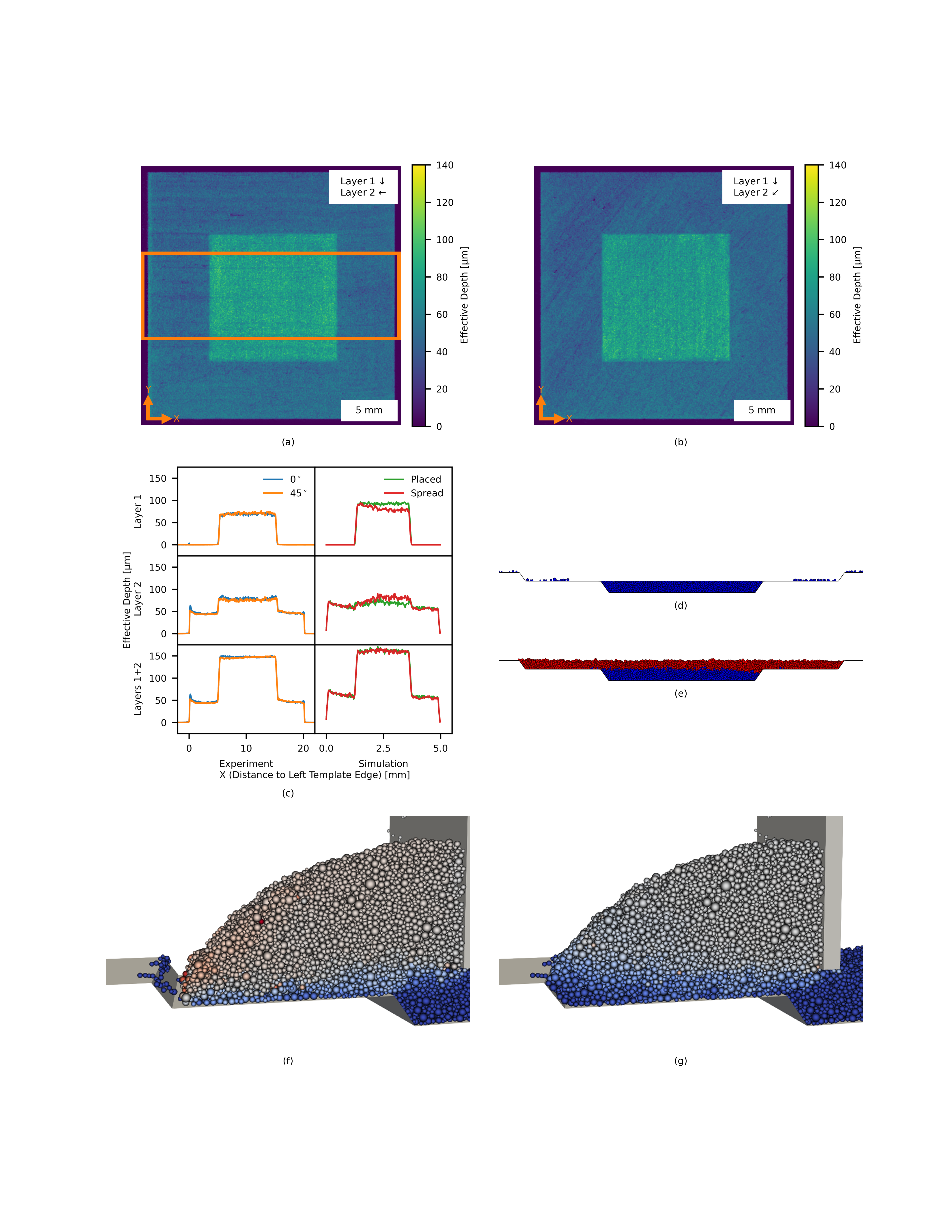}
	}
\end{center}
\caption{Multilevel spreading experiments and simulation.  (a) X-ray image of a multi-level spreading experiment (layers 1 + 2), wherein the left (X$=0$~mm) template edge is approached along the 0$^\circ$ direction (perpendicular to the first layer) when spreading the second layer.  (b) X-ray image of a multi-level recoating experiment, wherein the left template edge approached along the 45$^\circ$ direction.  (c) Average effective depth versus  coordinate, comparing experiments and simulations.  (d-e) Cross-sections of the simulated multi-layer experiment before and after simulating the second layer; layer 1 particles are blue and layer 2 particles red (see corresponding Supplementary Video~6).  (f-g) Particle velocities (blue-low, red-high) extracted from simulation before and after particles begin to interact with the left template edge, respectively (see corresponding Supplementary Video~5).}
\label{fig:Multilevel}
\end{figure*}

As described earlier, multi-level templates are fabricated to study how complex boundary conditions influence power layer packing density and uniformity. As such, Fig.~\ref{fig:Multilevel} illustrates use of a multi-level template with a first (perimeter) depth of 131~\textmu m and a second (center) depth of 306~\textmu m, or $175$~\textmu m below the perimeter, to enable study of powder deposition in the presence of sharp changes to local flow conditions.  Viewing the template from above, 15-45~\textmu m Ti-6Al-4V is first spread in the -Y direction (from the top to bottom with reference to the figure) to fill the lower (central) level of the template, excess is removed from the perimeter, and layer depth is determined by X-ray microscopy. The second layer is then spread into the upper (perimeter) region along the -X ($0^\circ$ direction), or at $45^\circ$ (upper right to lower left).  The total effective depth of both layers is then computed from a second transmission measurement, as illustrated for both spreading directions in Figs.~\ref{fig:Multilevel}a and~\ref{fig:Multilevel}b.  Critically, interpreting the attenuation measurements from the second layer requires two computed relationships of transmission versus effective-depth, as to account for additional beam hardening from the template in the shallow regions (due to the greater thickness of silicon template).  Finally, subtracting the second measurement from the first yields the effective depth of the second (upper) layer.  DEM simulations, as detailed in Section~\ref{sec:MethodDEM}, reproduce key geometric features of the experiment in a 5~mm long domain.  Two conditions are studied as to best compliment the experimental results.  The first considers spreading a second layer, in the -X direction per the convention above, over a first layer of particles directly placed into the deep region of the template.  These results are compared to the second configuration, in which both the first and second layers are created via spreading in the same direction.

Figure~\ref{fig:Multilevel}c presents effective depth versus X coordinate for each layer independently, as well as their sum.  For the experimental data, these curves are derived from averaging over the region indicated in orange in Fig.~\ref{fig:Multilevel}a.  It is first necessary to consider the increase in effective depth at distances proximal to the extreme left (X=0~mm) edge of the shallow region of the template, as is evident in the middle and bottom panels of Fig.~\ref{fig:Multilevel}c.  This increase in powder layer density begins several millimeters from the template edge, as powder particles at the front of the pile become trapped between the steep slope of the edge and the recoater blade.  Simulation of this geometry is shown in Figs.~\ref{fig:Multilevel}f and~\ref{fig:Multilevel}g, where particle color is indexed to velocity (see corresponding Supplementary Video~6).  We observe that particles near the edge in the latter figure are practically motionless and particles at the same depth but farther from the edge show leftward motion, but more slowly than the mean steady state pile velocity suggested by Fig.~\ref{fig:Multilevel}f.  Thus, particles adjacent to the edge become compressed into a more dense configuration as a consequence of this obstruction.  Experimental data show that deposition increases by 40\% (c.f. 21\% in simulation) at maximum if this change in boundary condition is approached directly at $0^\circ$.  Spreading the second layer at a $45^\circ$ angle to the template results only in a 24\% increase as the powder pile is somewhat able to flow along the template edge rather than be trapped by it.  Thus, features that substantially impede powder flow, such as protrusions from an underlying PBF AM part, may affect powder spreading dynamics over a large area. 

The second feature of interest lies in the effective depth of the upper (second) layer where it is spread over the deeper region of the template, as all experimental and simulated experiments agree that increased deposition occurs in this region as compared to the perimeter.  Two effects drive the observed increase.  First, the average surface of layer 1 lies slightly below the level of the shallow perimeter region, because particles above this level are pushed forward by the spreading blade.  Thus, particles spread as part of the second layer can settle into these gaps in low lying regions of the first layer.  Second, the DEM simulation shows that particles deposited in the first layer move due to forces exerted upon them when spreading the second layer.  Figure~\ref{fig:Multilevel}d, for example, shows a simulation with an initial configuration of layer 1 particles (blue), which are entrained in the subsequently spread powder (red) that forms the second layer in Fig.~\ref{fig:Multilevel}e (see corresponding Supplementary Video~5).  As the forces exerted by spreading are large enough to overcome friction and cause particle motion, they cause the layer to pack more densely.

We first address the simulation results in Fig.~\ref{fig:Multilevel}c, wherein density of the first layer is shown to impact the effective depth of the second.  In the top panel of simulation results, direct placement of the particles comprising layer 1 is shown to create an exceptionally dense and uniform initial configuration (green).  A packing fraction of approximately 0.52 ($\approx 91$~\textmu m effective depth in a $175$~\textmu m layer) is observed, approaching the simulated tap density of 0.58 and therefore leaving little room for densification via particle motion.  Thus, the effective depth of the second layer, depicted in the middle panel of simulation results in Fig.~\ref{fig:Multilevel}c, shows only a slight increase in effective depth from the perimeter to the central region ($\approx 56$~\textmu m to $\approx 69$~\textmu m).

Effective depth data from simulated spreading of layer 1 (red) must be interpreted with care due to the presence the same obstruction effect described above. Specifically, powder flow into the deep region while forming layer 1 is initially unimpeded, and therefore effective depth is constant at X coordinates spanning 2.5 to 3.75~mm, whereas increasing layer density is observed in the region 0.8$<$X$<$2.5~mm as the edge at X=0.8~mm is approached.  Considering the former portion with steady powder flow, a packing fraction of 0.45 is achieved.  An effective depth of $\approx 81$~\textmu m is observed when a second layer is spread above this area, substantially higher than the $\approx 69$~\textmu m observed from the placed particle case, indicating rearrangement of particles into a configuration of higher packing fraction.  Finally, we consider the region between X=0.8 and 2.5~mm in this simulated case.  The increase in layer 1 density in the vicinity of the template edge at X=0.8~mm is mirrored by decreased deposition in layer 2, as underlying material of increasingly higher density precludes further densification by spreading forces exerted in creating the second layer.  Thus, these results demonstrate that previous powder layers are subject to densification from disturbance forces arising during spreading of subsequent powder layers.

The experimental data strike the middle ground between the simulations.  No increase in powder deposition is observed at the left edges of the first layers (at X=5~mm), because of the choice of spreading directions and restriction of the area considered, and a packing fraction of 0.40 is observed.  Regardless of the spreading direction of the second layer, the effective depth is $\approx 80$ \textmu m over the central region, sharply increasing relative to $\approx 45$ \textmu m deposited in shallower perimeter region.  Gap filling alone is insufficient to explain this 35~\textmu m increase in effective depth over the central template region.  Figure~\ref{fig:Powder}a shows that 64\% of the particles in the size distribution lie below this dimension; therefore, there exist a sufficient number of particles in the distribution to make such a large gap statistically implausible.  Thus, again, a combination of gap filling and particle settling must be at play to explain the increase in powder deposition, and, in view of its substantial magnitude, is certain to couple to the physics of the fusion process (e.g., meltpool dynamics in LPBF).

\section{Conclusion}
\label{sec:conclusion}

Through spatially mapping powder layer density, this paper establishes X-ray microscopy as a high-fidelity technique for the study of powder spreading mechanics in AM.  Via experiments and supporting DEM simulations, we demonstrate relationships between relative layer thickness and powder cohesion on layer density, where thin layers and highly cohesive powders result in sparse deposition.  Defect morphology is also shown to depend on these variables, and provides preliminary guidance as to the role of stochastic powder layer variations on AM component quality.  Specifically, large powders cause correspondingly larger defects, as well as more total variation in effective depth at a given layer thickness. Abrupt changes to spreading boundary conditions are also studied, showing substantial alteration in powder deposition depending on the characteristics of the underlying material and at features that impede free powder flow.  

Future work will apply these methods in three distinct directions.  First, we envision study of powders themselves.  Tailored (e.g., skewed or bi-modal) size distributions could be used to increase packing density, and can be directly evaluated by X-ray microscopy.  Second, as highlighted in the introduction, optimal spreading tool designs and motion parameters remain an open question.  This imaging technique is compatible with in situ installation of a mechanized recoating testbed, and will enable matching of powder properties to recoater implements.  Along this direction, the uniformity of effective depth achieved by metered deposition or high applied shear (e.g., roller spreading mechanisms) for spreading highly cohesive powders can be studied.  Third, rigorous understanding of layer variation will guide selection of fusion process parameters, as to both mitigate defects arising from stochastic fluctuations in effective depth and provide for feedforward control in view of systematic, boundary-driven changes in powder deposition.  Finally, these investigations may be augmented via high-fidelity optical imaging of the spreading process and final layer, as to clarify the relations between powder rheology, particle motion in spreading, and ultimate material distribution.

\section*{Acknowledgements}
\label{sec:Acknowledgements}

Design and construction of the X-ray instrument, and experimentation, at MIT was supported by Honeywell Federal Manufacturing \& Technologies, Robert Bosch LLC, and Bayerische Motoren Werke AG (BMW).  R.W.P.\ was supported in part by a MathWorks Mechanical Engineering Fellowship at MIT. D.O.\ was supported by a NASA Space Technology Research Fellowship.  Work at TUM acknowledges funding by the Deutsche Forschungsgemeinschaft (DFG, German Research Foundation) within project 414180263.  We thank Kurt Broderick (MIT Microsystems Technology Laboratories) for providing extensive microfabrication expertise, Rachel Grodsky (Honeywell FM\&T) for performing laser diffraction (particle size) measurements, and Will Herbert (Carpenter Technology) for facilitating supply of metal powders.

\FloatBarrier

\bibliographystyle{elsarticle-num}
\bibliography{latex_main.bib}

\beginsupplement
\section{Silicon Template Fabrication}
\label{app:WaferFab}

Layer templates are fabricated from silicon wafers using standard microfabrication techniques.  Wafers are $100$~mm diameter, $1$~mm thick, intrinsic (undoped) silicon, oriented along the \textless100\textgreater{} crystal plane, and are supplied with a $2500$ \AA, stoichiometric LPCVD silicon nitride layer on both sides (Silicon Valley Microelectronics).  Fabrication begins by coating wafers with a negative photoresist (Futurrex NR9-1000PY).  The resist is subsequently patterned with a first photomask transparency (Fineline Imaging), delineating the deep regions of the multi-level templates, using an EV Group EVG620 mask aligner.  Resist development exposes the nitride layer in the aforementioned regions, enabling the unmasked nitride to be stripped via reactive ion etching (RIE) under tetrafluoromethane (System VII Plasmatherm).  Exposed silicon areas are etched using a 30 weight percent solution of potassium hydroxide in water, leveraging its high selectivity over silicon nitride~\cite{Williams2003}.  This solution is prepared on a hot/stir plate as to enable forced recirculation, critical to etch uniformity, and temperature is maintained at circa $65^\circ$C as to provide an etch rate of approximately $30$~\textmu m/min.  Etch rate is well approximated as exponential function temperature~\cite{Rai1997}; thus, wafers are frequently removed for measurement to assess the actual etch rate as the target dimensions are approached.  To fabricate the shallower regions of the templates, this process is repeated with a second photomask after an $0_2$ plasma cleaning to ensure photoresist adhesion.

\section{Angle of Repose Measurement}
\label{app:AoR}

\FloatBarrier

\FloatBarrier

\section{Simulation Videos}
\FloatBarrier
\begin{itemize}
    \item Video 1 - Perspective view of 15-45~\textmu m Ti-6Al-4V powder spread into a 131~\textmu m Profile A template.  Particles are colored by size.  Note particle motion is visible under the spreading blade.
    \item Video 2 -  Side view of the simulation in Video 1.
    \item Video 3 - Perspective view of 45-106~\textmu m Ti-6Al-4V powder spread into a 131~\textmu m Profile A template.  Particles are colored by size.  Note defects arising from large particles that cannot readily pass between the template bottom and spreading blade.
    \item Video 4 - Side view of the simulation in Video 3.
    \item Video 5 - Side view of a second layer of 15-45~\textmu m Ti-6Al-4V powder spread into a 131~\textmu m Profile B template.  Particles are colored by layer (layer 1 particles are blue, layer 2 particles are red).  Note the motion of blue particles as red particles are spread to form the second layer.
    \item Video 6 - Perspective view of a second layer of 15-45~\textmu m Ti-6Al-4V powder spread into a 131~\textmu m Profile B template.  Particles are colored by velocity.  Note the rapid deceleration of the particle pile as it interacts with the template edge.
\end{itemize}
\FloatBarrier
\end{document}